\documentclass[preprint,12pt]{elsarticle}

\usepackage{lineno,hyperref,graphicx}

\usepackage[T1]{fontenc}
\usepackage{amsmath,upgreek}
\usepackage{amssymb}
\usepackage{url}
\usepackage[utf8]{inputenc}
\DeclareUnicodeCharacter{2014}{\dash}
\DeclareUnicodeCharacter{2212}{-}

\newcommand{\pd}[3]{\frac{\partial^{#3} #1}{\partial {#2}^{#3}}} 
\newcommand{\td}[3]{\frac{\mathrm{d}^{#3} #1}{\mathrm{d} {#2}^{#3}}} 

\journal{Astroparticle physics}

\begin{document}
\begin{frontmatter}

\title{Connecting multi-lepton anomalies at the LHC and in Astrophysics with MeerKAT/SKA}

\author[1]{Geoff Beck\corref{cor1}}
\ead{geoffrey.beck@wits.ac.za}

\author[1]{Ralekete Temo}
\ead{1497117@students.wits.ac.za}

\author[2]{Elias Malwa}
\ead{elias.malwa@cern.ch}

\author[2]{Mukesh Kumar}
\ead{mukesh.kumar@wits.ac.za}

\author[2,3]{Bruce Mellado}
\ead{bruce.mellado@wits.ac.za}

\cortext[cor1]{Corresponding author}

\address[1]{School of Physics and Centre for Astrophysics, University of the Witwatersrand, Wits 2050, Johannesburg, South Africa}
            
\address[2]{School of Physics and Institute for Collider Particle Physics, University of the Witwatersrand, Wits 2050, Johannesburg, South Africa}
            
\address[3]{iThemba LABS, National Research Foundation, PO Box 722, Somerset West, 7129, South Africa}

\begin{abstract}
Multi-lepton anomalies at the Large Hadron Collider (LHC) are reasonably well described by a two Higgs doublet model with an additional singlet scalar. Here we demonstrate that using this model, with parameters set by the LHC, we are also able to describe the excesses in gamma-ray flux from the galactic centre and the cosmic-ray spectra from AMS-02. This is achieved through Dark Matter (DM) annihilation via the singlet scalar. Of great interest is the flux of synchrotron emissions which results from annihilation of DM in Milky-Way satellites. We make predictions for MeerKAT/SKA observations of the nearby dwarf galaxy Reticulum II and we demonstrate the power of this instrument as a new frontier in indirect dark matter searches. Since the dark matter sector of the aforementioned two Higgs doublet model is unconstrained by current LHC data, we also demonstrate a synergy between particle and astrophysical searches in order to motivate further exploration of this promising model.
\end{abstract}

\begin{highlights}
\item Dark matter candidate drawn from LHC data
\item Indirect detection limits on LHC inspired dark matter
\end{highlights}

\end{frontmatter}

\section{Introduction}
The discovery of a Higgs boson ($h$) \cite{Higgs:1964ia,Englert:1964et,Higgs:1964pj,Guralnik:1964eu} at the Large Hadron Collider (LHC) by the
ATLAS \cite{Aad:2012tfa} and CMS \cite{Chatrchyan:2012ufa} experiments has opened a new chapter in particle physics. Measurements provided 
so far indicate that the quantum numbers of this boson are consistent with those predicted by the Standard Model (SM) \cite{Chatrchyan:2012jja,Aad:2013xqa}, 
and that the relative branching ratios (BRs) to SM particles are similarly well described. With this in mind, a window of opportunity now opens for the search for new bosons.

In this paper we will be exploring the possibility of a two-Higgs doublet model with an additional singlet scalar $S$ (2HDM+$S$) providing a Dark Matter (DM) candidate. 
One of the previously studied implications of a 2HDM+$S$ model is the production of multiple-leptons through the decay chain $H\rightarrow Sh,SS$ \cite{vonBuddenbrock:2016rmr}, where $H$ is the heavy CP-even scalar and $h$ is considered as the SM Higgs boson with mass $m_h = 125$ GeV. Excesses in multi-lepton final states were reported in \cite{vonBuddenbrock:2017gvy}. In order to further explore results with more data and new final states, while avoiding biases and look-else-where effects, the parameters of the model were fixed in 2017 according to \cite{vonBuddenbrock:2016rmr,vonBuddenbrock:2017gvy}. This includes setting the scalar masses as $m_H=270$ GeV, $m_S=150$ GeV, treating $S$ as a SM Higgs-like scalar and assuming the dominance of the decays $H\rightarrow Sh,SS$. Statistically compelling excesses in opposite sign di-leptons, same-sign di-leptons, and three leptons, with and without the presence of $b$-tagged hadronic jets were reported in \cite{vonBuddenbrock:2019ajh,vonBuddenbrock:2020ter,Hernandez:2019geu}. Subsequently, with independent data sets and the phase-space fixed by the above mentioned model, an 8 $\sigma$ combined excess was reported in \cite{Buddenbrock:2019tua}, indicating a statistical preference for 2HDM+$S$ over the SM alone. These excesses have continued to grow \cite{Hernandez:2019geu,vonBuddenbrock:2020ter,Crivellin:2021ubm,Fischer:2021sqw}.\footnote{For a recent review of the multi-lepton anomalies see plenary talk at ZPW2023, January 12-14, Zurich,  https://www.physik.uzh.ch/events/ZPW2023/.} With the procedure followed, the look-else-where or trials factors are nullified. In addition, a candidate of $S$ with a mass of 151.5 GeV and a global significance of 3.9$\sigma$ has just been reported in \cite{Crivellin:2021ubm}.
The possible connection with the anomalous magnetic moment of the muon $g-2$ was reported in \cite{Sabatta:2019nfg}. For a review of various anomalies see \cite{Fischer:2021sqw}. Other authors have suggested alternative explanation of the multi-lepton anomalies, e.g., \cite{Jueid:2021avn,Kanemura:2021dez,Fischer:2021nha,Antusch:2021oit,Aboubrahim:2021phn}.

Interestingly, the aforementioned 2HDM+$S$ models can accommodate a DM candidate particle, whose production would contribute to ``missing energy" in LHC measurements. As yet, the properties of this particle are not constrained by any of the data from the LHC used in \cite{vonBuddenbrock:2016rmr,vonBuddenbrock:2017gvy,vonBuddenbrock:2019ajh,vonBuddenbrock:2020ter,Hernandez:2019geu}. This can be utilised to further test the 2HDM+$S$ model in the context of astrophysical measurements, which indicate that baryonic matter comprises 5\% of the current energy density in the universe while DM makes up more than 24\% \cite{2020Planck,Koopmans_2003,Metcalf_2004,Hoekstra_2002,Moustakas_2003}. Thus, the DM candidate of the model could be expected to be probed by observations of regions with high DM density.  

In this study we aim to use astrophysics as an indirect probe of the unconstrained elements of the 2HDM+$S$ model that is, so far, motivated by LHC anomalies. This is done via a DM particle coupling to $S$ as a mediator to the SM, where the non-DM parameters of the collider model are fixed to describe the LHC data \cite{vonBuddenbrock:2016rmr,vonBuddenbrock:2018xar}. Since the DM candidate is unconstrained at the LHC, additional observational probes are needed to supplement the collider data. Thus, we will study the potential DM candidate parameter space while holding the other 2HDM+$S$ parameters fixed. We make particular use of the observed positron excess by the Alpha Magnetic Spectrometer (AMS-02) \cite{PhysRevLett.122.041102}, anti-protons from the same detector \cite{Aguilar:2016kjl}, and the excess in gamma-ray fluxes from the galactic centre measured by Fermi-LAT \cite{Ackermann_2017}. This is partially motivated by the fact that the LHC anomalies triggering the 2HDM+$S$ model are leptonic in nature. These particular astrophysical data sets are of interest as they have been extensively studied as potential signatures of DM \cite{Cholis_2019,amse1,amse2,amse3,Profumo:2019pob,amse5,amsp1,Beck_2019,gs2016}. This DM model is then used to make predictions for radio observations with the MeerKAT precursor to the Square Kilometre Array (SKA). These emissions would result from synchrotron radiation from electrons and positrons produced in DM annihilations. As such, we compute the number of $e^{+}$ and $p^{-}$ as a result of DM annihilations using the model described here from the collider physics perspective, and with the input from the astrophysics, we extend this by ensuring consistency with astrophysical observations. With a constrained DM model, our predictions can then be tested independently via observations with the MeerKAT telescope. In this regard, the complementarity between collider and astroparticle physics is investigated in the model considered here. Furthermore, the favoured DM candidate mass regions, resulting from astrophysical limits, can be used to motivate collider searches to better probe the properties of the 2HDM+$S$ model.

This paper is structured as follows. The particle and astro-particle physics models are described in Sections~\ref{sec:model} and~\ref{sec:astromodel}, respectively. Then, Section~\ref{sec:meerkat} succinctly describes the MeerKAT telescope. The methodology used here is detailed in Section~\ref{sec:methodology}. Results and conclusions are presented in Sections~\ref{sec:results} and~\ref{sec:conclusions} respectively.

\section{Particle physics model}
\label{sec:model}
Here, we succinctly describe the model used to describe the multi-lepton anomalies observed in the LHC data and with which to interpret the above mentioned excesses in astrophysics. The formalism is comprised of a model of fundamental interactions interfaced with a model of cosmic-ray fluxes that emerge from DM annihilation.
The potential for a two Higgs-doublet model with an additional real singlet field $\Phi_S$ (2HDM+$S$) is given as in \cite{vonBuddenbrock:2016rmr}:
\begin{align}
V(\Phi)
=&\, m^2_{11}\left|\Phi_{1}\right|^2 + m^2_{22}\left|\Phi_{2}\right|^2 - m^2_{12}(\Phi^\dagger_{1}\Phi_{2}+ {\rm h.c.}) \notag \\
&+\frac{\lambda_{1}}{2}\left(\Phi^{\dagger}_{1}\Phi_{1}\right)^2  +\frac{\lambda_{2}}{2}\left(\Phi^{\dagger}_{2}\Phi_{2}\right)^2 +\lambda_{3}\left(\Phi^\dagger_{1}\Phi_{1}\right)\left(\Phi^\dagger_{2}\Phi_{2}\right) \notag\\ 
& + \lambda_{4}\left(\Phi^\dagger_{1}\Phi_{2}\right)\left(\Phi^\dagger_{2}\Phi_{1}\right)
+\frac{\lambda_{5}}{2}\Big[\left(\Phi^\dagger_{1}\Phi_{2}\right)^2+ \rm{h.c.}\Big] \notag\\
& + \frac{1}{2} m_S^2 \Phi_S^2 + \frac{\lambda_{6}}{8}\Phi^4_{S}  + \frac{\lambda_{7}}{2}\left(\Phi^\dagger_{1}\Phi_{1}\right)\Phi^2_{S} \notag \\ & + \frac{\lambda_{8}}{2}\left(\Phi^\dagger_{2}\Phi_{2}\right)\Phi^2_{S}.
\label{pot}
\end{align}
The fields $\Phi_1$, $\Phi_2$ in the potential are the $SU(2)_L$ Higgs doublets. The first three lines in Eq.~(\ref{pot}) are the contributions of the real 2HDM potential. The terms of the last line are contributions of the singlet field $\Phi_S$. In order to prevent tree-level flavour-changing neutral currents (FCNCs), we extend Yukawa's sector with $\mathbb{Z}_2$ symmetry, which has a trivial solution under the following transformation:
\begin{align}
    \Phi_1 \rightarrow \Phi_1 ,\quad \Phi_2 \rightarrow -\Phi_2 ,\quad \Phi_S \rightarrow  \Phi_S.
\end{align}
In addition, we can consider another $\mathbb{Z}^\prime_2$ symmetry of the transformation:
\begin{align}
     \Phi_1 \rightarrow \Phi_1 ,\quad \Phi_2 \rightarrow \Phi_2 ,\quad \Phi_S \rightarrow -\Phi_S.
\end{align}
Having imposed a $\mathbb{Z}_{2}$ symmetry for the transformations of the form $ h \to h, H \to -H$ and
$S \to S$, the $\lambda_3$ term will vanish which also eliminates $\lambda_6$ and $\lambda_7$ from $V(\phi_1, \phi_2)$. However, we assume a soft breaking of $\mathbb{Z}^\prime_{2}$ symmetry and the following fields: $\Phi_{1}\to \frac{v1}{\sqrt{2}},\Phi_{2}\to \frac{v2}{\sqrt{2}}$ and $\Phi_{S}\to v_{S}$  acquire $vev$ after EWSB which allows $m^2_{12} = 0$. After spontaneous EWSB, five physical Higgs particles are left in the spectrum: one charged Higgs pair, $H^{\pm}$, one
$CP$-odd scalar, $A$, and two $CP$-even states, $h$ and $H$. We impose two important angles ($\alpha$ and $\beta$) which constrain the model parameter spaces. The angle $\alpha$ diagonalises the $CP$-even Higgs squared-
mass matrix and $\beta$ diagonalises both the $CP$-odd and charged Higgs sectors, which leads to $\tan \beta = \frac{v_2}{v_1}$.  By fixing the lighter Higgs mass, $m_h = 125.09$ GeV, $CP$ odd scalar mass $m_A = m_H + 100$ GeV and $CP$ even scalar mass $m_{H^{\pm}} = m_H + 100$ GeV, the heavy Higgs mass $m_H$ and $\tan \beta$ is scanned. The model parameter space is constrained such that $\cos (\beta - \alpha) \leq 0.5(0.2)$, $m_H \leq 380(\approx 380)$ and $\tan \beta \leq 2$.
After the minimisation of the potential and electro-weak symmetry breaking, the scalar sector is populated with three $CP$ even scalars $h$ (SM Higgs), $H$ and $S$, one $CP$ odd scalar $A$ and charged scalar $H^\pm$. These constraints have been obtained by considering the decay channels in Ref~\cite{CMS:2014suk,CMS:2015hra,CMS:2015mba}.
Minimizing the potential in Eqn.1~, the following mass eigenstates of $v_{1}, v_{2}$ and $v_{S}$ are obtained:
\\
\begin{align}
    m^2_{11} = -\frac{1}{2}(v^2_{1}\lambda_1 + v^2_2\lambda_{7}) + \frac{v_{2}}{v_{1}}m^2_{12}),\\
    m^2_{22} = -\frac{1}{2}(v^2_{2}\lambda_2 + v^2_{1}\lambda_{345} + v^2_{S}\lambda_{8} + \frac{v_2}{v_1}m^2_{12},\\
    m^2_S = -\frac{1}{2}(v^2_1\lambda_{7} + v^2_{2}\lambda_8 + v^2_S\lambda_{6}),
\end{align}
where the scalar sector is populated with three $CP$ even scalars $h, H$ and $S$, one $CP$ odd scalar $A$ and charged scalar $H^\pm$. Following are the production modes for $S$,
i.e $S \to b \bar{b},\tau^{+}\tau^{-}, \mu^{+}\mu^{-}, s\bar{s}, c\bar{c}, gg, \gamma\gamma, Z\gamma,VV (V=Z,W^{\pm}$) whereby $S$ can be located in one of two prominent locations. The first is dominated by $S \to VV (V=Z,W^{\pm}$) when $m_S \geq 2m_W \approx 160$~GeV. The second is when $m_S \leq 2m_W$, and in this region $S$ has non-negligible BRs to various decay products such as $b\bar{b},VV gg, \gamma\gamma, Z\gamma$ etc. 
More details of this model and associated interactions' Lagrangians and parameter space we refer to  \cite{vonBuddenbrock:2016rmr,vonBuddenbrock:2018xar}. 
Further, we consider interactions of $S$ with three types of DM candidate ($\chi$). These will correspond to the possibility of $\chi$ having either spin 0, 1/2, or 1. The interactions take the forms: 
\begin{align}
\mathcal{L}_{\mathrm{int},0} & = \frac{1}{2} m_{\chi} g^{S}_{\chi}\chi\chi S \; , \\
\mathcal{L}_{\mathrm{int},1/2} & = {\bar{\chi}} (g^S_{\chi}+ig^P_{\chi}\gamma_{5})\chi S \; , \\ \mathcal{L}_{\mathrm{int},1} & = g^S_{\chi}\chi^{\mu}\chi_{\mu}S \; , \label{intdm}
\end{align}
where $g^S_{\chi}$ is the strength of the scalar coupling between DM and the singlet real scalar $S$, $g^P_\chi$ is the strength of the pseudo-scalar coupling to $S$, and $m_{\chi}$ is the DM mass. Having these interactions in mind, we consider the processes $\chi \, \bar{\chi} \to S \to X$ and $\chi \, \bar{\chi} \to S \to H \, S/h \to X$ (where $X$ represents some Standard Model products). We will term these 2$\rightarrow$2 and 2$\rightarrow$3 for brevity. The motivation for this pair of interactions is that they are the simplest options available. 

It is important to note that the non-DM model parameters for the 2HDM+$S$ model are fixed according to \cite{vonBuddenbrock:2016rmr,vonBuddenbrock:2017gvy}. Some of the most relevant values here being that $m_S = 150$ GeV and $m_H = 270$ GeV. Since the $S$ boson is assumed to have Higgs-like branching ratios to the SM, the mass of these particles will directly effect the $\chi\chi \to$ SM products, as the SM branching ratios scale with $m_S$. 

Finally, the DM candidate degree of freedom is, as yet, unconstrained by LHC data. Therefore, we will scan a wide parameter space, from the lowest DM masses that can produce real $S$ bosons up to 1 TeV. In the $2\to3$ case, the lower mass limit is that required to produce $H$ and $h$ bosons together. In this regard, astrophysical searches will compliment LHC data by probing aspects of the 2HDM+$S$ model that are not yet determined by collider data.

\section{Astrophysical modelling}
\label{sec:astromodel}
Here we will discuss the models used to determine both cosmic-ray and photon fluxes arriving at Earth as a result of DM annihilation. Considering the source halos as the Milky Way (cosmic rays) and Reticulum II (photons).

The principle equation of interest, in this regard, will be the diffusion-loss equation for a particle species $i$:
\begin{equation}
\pd{}{t}{} \td{n_{i}}{E}{} = \vec{\nabla}\cdot\left(D(E,\vec{x})\vec{\nabla}\td{n_{i}}{E}{}\right) + \pd{}{E}{}\left[b(E,\vec{x})\td{n_{i}}{E}{}\right] + Q_{i}(E,\vec{x}) \; , \label{eq:difusion-eq}
\end{equation}
where $D$ is the diffusion function, $b$ is the energy-loss function, $Q_{i}$ is the source function for $i$-particles, and $\td{n_{i}}{E}{}= \td{n_{i}}{E}{}(E,\vec{x})$ is the number density per unit energy of $i$-particles. In particular, the equilibrium solutions of Eq.~(\ref{eq:difusion-eq}) will provide the distributions of particles injected by DM annihilations within a given astrophysical environment.

In all considered scenarios we will require a source function. This represents the rate of injection of DM annihilation products $i$, with energy $E$, into the halo at position $\vec{x}$. It is given by 
\begin{equation}
Q_{i}=\frac{1}{2}\left(\frac{\rho_\chi(\vec{x})}{m_{\chi}}\right)^2{\langle \sigma{V} \rangle} \left. \td{n_{i}}{E}{}\right\rvert_{\mathrm{inj}} \; , \label{eq:source}
\end{equation}
where $\left. \td{n_{i}}{E}{}\right\rvert_{\mathrm{inj}}$ is the injected number density of $i$-particles per unit energy, $\rho_\chi(\vec{x})$ is the DM density at $\vec{x}$, and ${\langle \sigma{V} \rangle}$ is the velocity averaged annihilation cross-section for DM particles.

\subsection{Cosmic rays fluxes}
Here we will outline the method for calculating the fluxes of positrons and anti-protons that arrive at Earth as a result of DM annihilation within the Milky Way galactic centre. To compute these fluxes we will rely upon equilibrium solutions $\td{n_{i}}{E}{}$ to Eq.~(\ref{eq:difusion-eq}) taken from \cite{ppdmcb1}. In doing so we note that the function $D(E,\vec{x})$ depends upon assumptions about the Milky-Way diffusion environment \cite{maurin2001cosmic} and we explore all three value sets MIN, MED, and MAX presented in \cite{ppdmcb1}. Note that these diffusion scenarios have been updated in \cite{G_nolini_2021}. The main difference to the current case would be that that DM-induced fluxes are slightly increased for all three scenarios~\cite{G_nolini_2021}. However, at present no suitable propagation functions are available for use.

For positrons, these equilibrium solutions yield the flux in the solar neighbourhood via:
\begin{equation}
\label{eq:4}
\td{\Phi_{e^+}}{E}{}= \;\frac{c {\langle \sigma{V} \rangle}}{8{\pi}b(E)}\left(\frac{\rho_\odot}{m_{\chi}}\right)^2 \int_{E}^{m_{\chi}} dE_{s}\left. \td{n_{e^+}}{E_s}{}\right\rvert_{\mathrm{inj}}I_\odot(E,E_s) \; ,
\end{equation}
where $E_s$ is the energy of injected positrons, and $I_\odot(E,E_s)$ is a Green's function solving Eq.~(\ref{eq:difusion-eq}) at the location of Earth, this being given by \cite{ppdmcb1}. 

Similarly, the anti-proton flux in the solar neighbourhood can be determined, according to \cite{ppdmcb1}, as being given by:
\begin{equation}
    \td{\Phi_{\bar{p}}}{K}{} = \frac{v_{\bar{p}}}{8\pi} {\langle \sigma{V} \rangle} \left(\frac{\rho_\odot}{m_{\chi}}\right)^2 R(K) \langle \sigma V\rangle \left. \td{n_{\bar{p}}}{K}{}\right\rvert_{\mathrm{inj}} \; , \label{eq:antiprotons}
\end{equation}
where $K$ is anti-proton kinetic energy, $v_{\bar{p}}$ is the anti-proton speed, $R(K)$ are the propagation functions from \cite{ppdmcb1}, and $\left. \td{n_{\bar{p}}}{K}{}\right\rvert_{\mathrm{inj}}$ is the injected anti-proton spectrum.

\subsection{Photon fluxes}
\label{sec:photons}
This section will be concerned with photons produced within the environment of the Reticulum II dwarf galaxy. These include primary photons, produced in the annihilation of DM, and secondary photons that result from the interaction of annihilation products (electrons and positrons) with their environment. These latter mechanisms will be inverse-Compton scattering of CMB photons, bremsstrahlung with ambient ions, and synchrotron radiation due to the magnetic field within Reticulum II itself.

\subsubsection{Primary fluxes}
Primary photon fluxes within radius $r$ of a halo centre, at frequency $\nu$ are found via:
\begin{equation}
S_{\gamma} (\nu,r) = \int_0^r d^3r^{\prime} \, \frac{Q_{\gamma}(\nu,r^\prime)}{4\pi (d_L^2+\left(r^\prime\right)^2)} \; , \label{eq:gamma}
\end{equation} 
where $Q_{\gamma}$ is the photon source function and where $d_L$ is the luminosity distance to the halo centre.  The integral in Eq.~(\ref{eq:gamma}) will be taken over some chosen region of interest in each studied target.

\subsubsection{Secondary fluxes}
Secondary photon fluxes (from electrons produced by DM annihilation) are found using the emissivity:
\begin{equation}
j_{i} (\nu,r) = \int_{0}^{m_{\chi}} dE \, \td{n_{e^\pm}}{E}{}(E,r) P_{i} (\nu,E,r) \; ,
\label{eq:emm-he}
\end{equation}
where $\td{n_{e^\pm}}{E}{}$ is the sum of the electron and positron distributions (equilibrium solutions to Eq.~(\ref{eq:difusion-eq})) within the source region and $P_i$ is the power emitted at frequency $\nu$ through mechanism $i$ by an electron with energy $E$, at position $r$. The $P_i$ functions are detailed later. The flux produced within a radius $r$ is then found via:
\begin{equation}
S_{i} (\nu,r) = \int_0^r d^3r^{\prime} \, \frac{j_{i}(\nu,r^{\prime})}{4 \pi (d_L^2+\left(r^{\prime}\right)^2)} \; .
\label{eq:flux}
\end{equation}

As the environment we consider is one of a dwarf galaxy we must take diffusion into account when solving Eq.~(\ref{eq:difusion-eq}) \cite{baltz2004} to determine $\td{n_{e^\pm}}{E}{}$. For all secondary fluxes we solve Eq.~(\ref{eq:difusion-eq}) with a Green's function method. Assuming spherical symmetry, we have that \cite{baltz1999,baltz2004,Colafrancesco2006,Colafrancesco2007}:
\begin{equation}
\td{n_e}{E}{} (r,E) = \frac{1}{b(E)}  \int_E^{m_{\chi}} d E^{\prime} \, G(r,E,E^{\prime}) Q_e (r,E^{\prime}) \; ,
\end{equation}
this solution method requires that $b$ and $D$ have no spatial dependency, thus they are defined
\begin{equation}
D(E) = D_0 \left(\frac{d_0}{1 \; \mbox{kpc}}\right)^{\frac{2}{3}} \left(\frac{\overline{B}}{1 \; \mu\mbox{G}}\right)^{-\frac{1}{3}} \left(\frac{E}{1 \; \mbox{GeV}}\right)^{\frac{1}{3}}  \; , \label{eq:diff}
\end{equation}
where $D_0 = 3.1\times 10^{28}$ cm$^2$ s$^{-1}$ \cite{Regis_2014}, $d_0$ is the magnetic field coherence length, $\overline{B}$ is the average magnetic field strength, and $E$ is the electron energy. The loss-function is found via \cite{Colafrancesco2006,egorov2013}:
\begin{equation}
\begin{aligned}
b(E) & = b_{\mathrm{IC}} \left(\frac{E}{1\,\mathrm{GeV}}\right)^2 + b_{\mathrm{sync}} \left(\frac{E}{1\,\mathrm{GeV}}\right)^2 \left(\frac{\overline{B}}{1 \; \mu\mbox{G}}\right)^2 \; \\ & + b_{\mathrm{Coul}} \left(\frac{\overline{n}}{1 \; \mbox{cm}^{-3}}\right) \left(1 + \frac{1}{75}\log\left(\frac{\gamma}{\left(\frac{\overline{n}}{1 \; \mbox{cm}^{-3}}\right)}\right)\right) \\& + b_{\mathrm{brem}} \left(\frac{\overline{n}}{1 \; \mbox{cm}^{-3}}\right) \left(\frac{E}{1\,\mathrm{GeV}}\right)\;,
\end{aligned}
\label{eq:loss}
\end{equation}
where $\overline{n}$ is the average gas density, the coefficients $b_{\mathrm{IC}}$, $b_{\mathrm{sync}}$, $b_{\mathrm{Coul}}$, $b_{\mathrm{brem}}$ are the energy-loss rates from ICS, synchrotron emission, Coulomb scattering, and bremsstrahlung. These coefficients are given by $0.25\times 10^{-16}(1+z)^4$ (for CMB target photons), $0.0254\times 10^{-16}$, $6.13\times 10^{-16}$, $4.7\times 10^{-16}$ in units of GeV s$^{-1}$. The average quantities $\bar{n}$ and $\bar{B}$ are computed by weighting the average with $\rho_\chi^2$, ensuring they accurately reflect the environment of the majority of annihilations. 

The Green's function is then given by:
\begin{align}
G(r,E,E^{\prime}) = & \frac{1}{\sqrt{4\pi\Delta v}} \sum_{n=-\infty}^{\infty} (-1)^n \int_0^{r_{\mathrm{max}}} d r^{\prime} \; \frac{r^{\prime}}{r_n} f_{G,n}(r,r^\prime,\Delta v) \; , \\ 
f_{G,n}(r,r^\prime,\Delta v) & =  \left( \mathrm{e}^{\left(-\frac{\left(r^{\prime} - r_n\right)^2}{4\Delta v}\right)} - \mathrm{e}^{\left(-\frac{\left(r^{\prime} + r_n\right)^2}{4\Delta v}\right)} \right)\frac{Q_e(r^{\prime})}{Q_e(r)} \; ,
\end{align}
with the sum running over a set of image charges, each at position $r_n = (-1)^n r + 2 n r_{\mathrm{max}}$, where $r_{\mathrm{max}} = 2 r_{\mathrm{vir}}$ or twice the virial radius. The additional required definitions have:
\begin{equation}
\Delta v =  v(u(E)) - v(u(E^{\prime})) \; ,
\end{equation}
with
\begin{equation}
\begin{aligned}
v(u(E)) = & \int_{u_{\mathrm{min}}}^{u(E)} dx \; D(x) \; , \\
u (E) = & \int_E^{E_{\mathrm{max}}} \frac{dx}{b(x)} \; , \\ 
\end{aligned}
\end{equation}
where $E_{\mathrm{max}} = m_{\chi}$.

\subsubsection{Secondary emission power functions}
In this section we will detail the power functions $P_i$ for synchrotron, inverse-Compton, and bremsstrahlung emission. 

For synchrotron emission we have that an electron of energy $E$, at position $r$ radiates power at frequency $\nu$ given by \cite{longair1994,rybicki1986}:
\begin{equation}
P_{\mathrm{sync}} (\nu,E,r) = \int_0^\pi d\theta \, \frac{\sin{\theta}^2}{2}2\pi \sqrt{3} r_e m_e c \nu_g  F_{\mathrm{sync}}\left(\frac{\kappa}{\sin{\theta}}\right)  \; ,
\label{eq:power}
\end{equation}
where $\nu_g$ is the non-relativistic gyro-frequency, $r_e$ is the electron radius, $m_e$ is the electron mass, and $\theta$ is the angle between the magnetic field and electron trajectory. The value of $\kappa$ is found via:
\begin{equation}
\kappa = \frac{2\nu}{3\nu_g \gamma^2}\left[1 +\left(\frac{\gamma \nu_p}{\nu}\right)^2\right]^{\frac{3}{2}} \; ,
\end{equation}
with $\gamma = \frac{E}{m_e c^2}$. Finally,
\begin{equation}
F_{\mathrm{sync}}(x) \simeq 1.25 x^{\frac{1}{3}} \mbox{e}^{-x} \left(648 + x^2\right)^{\frac{1}{12}} \; .
\end{equation}

The power produced by the inverse-Compton scattering (ICS) at a photon of frequency $\nu$ from an electron with energy $E$ is given by \cite{longair1994,rybicki1986}:
\begin{equation}
P_{\mathrm{IC}} (\nu,E) = c E_{\gamma}(z) \int d\epsilon \; n(\epsilon) \sigma(E,\epsilon,E_{\gamma}) \; ,
\label{eq:ics_power}
\end{equation}
where $\epsilon$ is the energy of the seed photons distributed according to $n(\epsilon)$ (this will taken to be that of the CMB), and
\begin{equation}
\sigma(E,\epsilon,E_{\gamma}) = \frac{3\sigma_T}{4\epsilon\gamma^2}G(q,\Gamma_e) \; ,
\end{equation}
with $\sigma_T$ being the Thompson cross-section and
\begin{equation}
G(q,\Gamma_e) = 2 q \ln{q} + (1+2 q)(1-q) + \frac{(\Gamma_e q)^2(1-q)}{2(1+\Gamma_e q)} \; ,
\end{equation}
with
\begin{equation}
\begin{aligned}
q & = \frac{E_{\gamma}}{\Gamma_e(\gamma m_e c^2 + E_{\gamma})}, \quad
\Gamma_e & = \frac{4\epsilon\gamma}{m_e c^2} \; ,
\end{aligned}
\end{equation}
where $m_e$ is the electron mass.

Finally, the power from bremsstrahlung at photon energy $E_\gamma$ from an electron at energy $E$ is given by \cite{longair1994,rybicki1986}:
\begin{equation}
P_{\mathrm{brem}} (E_{\gamma},E,r) = c E_{\gamma}\sum\limits_{j} n_j(r) \sigma_B (E_{\gamma},E) \; ,
\end{equation}
where $n_j$ is the distribution of target nuclei of species $j$ and the cross-section is given by:
\begin{equation}
\sigma_B (E_{\gamma},E) 
=\frac{3\alpha \sigma_T}{8\pi E_{\gamma}}\left[ \left(1+\left(1-\frac{E_{\gamma}}{E}\right)^2\right)\phi_1 - \frac{2}{3}\left(1-\frac{E_{\gamma}}{E}\right)\phi_2 \right],
\end{equation}
with $\phi_1$ and $\phi_2$ being energy dependent factors determined by the species $j$ (see \cite{longair1994,rybicki1986}).

\section{MeerKAT and the SKA}
\label{sec:meerkat}
Indirect detection of DM has been traditionally focused upon the use of gamma-ray experiments, such as the Fermi-LAT \cite{atwood2009large}, because this mode of detection has low attenuation in the interstellar medium and has high detection efficiency. Recently, the indirect hunt for DM in radio-band has become prominent. This emerges from the fact that radio interferometers have an angular resolution vastly transcending that of gamma-ray experiments. 
Of particular interest to this work is the SKA, an international science project designed for studies in the field of radio astronomy. This telescope array provides around 50 times the sensitivity and 10,000 times the survey speed of the best current telescopes \cite{ska2012}. These capabilities have already been extensively argued to provide a powerful tool for exploring the properties of DM via indirect detection of annihilation or decay products \cite{gsp2015,gs2016,Beck_2019}. At present, the precursor array MeerKAT is currently being operated by the South African Radio Astronomy Observatory (SARAO) with 64 antennae elements. Each of the elements is a 13.5 m diameter dish, configured to achieve high sensitivity and wide field-of-view imaging of the sky. With 20 hours of time on target it is estimated that MeerKAT can achieve an rms sensitivity of $2.7$ $\mu$Jy beam$^{-1}$ at robust weighting $0$, this is sourced from SARAO's publicly available tools\footnote{\url{https://apps.sarao.ac.za/calculators/continuum}}. This results in sensitivity, at arcminute scales, of $= 2.7\, \mu\mathrm{Jy} \, \mathrm{beam}^{-1} \times \left(\frac{1}{8\,\mathrm{arcseconds} \, \mathrm{beam}^{-1}}\right)^2 \frac{\pi}{4\log(2)} \approx 172$ $\mu$Jy arcminute$^{-2}$, when taking into account a gaussian synthesized beam size of $\approx 8$ arcseconds. We note that a more tailored estimate of MeerKAT sensitivity, making use of an arcminute-scale taper on the visibilities, will likely yield an improved sensitivity. This is because the DM emission is diffuse and will be on a scale of arcminutes \cite{regis2017}, while the visibility taper reduces contributions from long array baselines \cite{radioAstronFund}, corresponding to small angular scales, which would be dominated by the signal from point-sources. Therefore, $172$ $\mu$Jy therefore constitutes a conservative estimate for arcminute scale sensitivity to diffuse emission. Note that the impact of the taper would be somewhat reduced by the additional need for the subtraction of non-DM emission sources.
 
The sensitivity of MeerKAT is around a factor of 2 better than ATCA\footnote{\url{https://www.narrabri.atnf.csiro.au/myatca/interactive_senscalc.html}} which has previously been used for indirect DM searches in dwarf galaxies \cite{regis2017}. This notable sensitivity advantage is a consequence of the instrument exceeding its original design specifications, see \cite{booth2009meerkat,ska2012}, making it an unexpected new leader in radio-frequency DM searches. Additionally, MeerKAT is expected to receive an upgrade of an additional 13 dishes, taking it to 77, with construction expected to be complete in 2023.\footnote{\url{https://www.mpg.de/15382572/top-radio-telescope-in-south-africa}}

When computing SKA sensitivities we make use of Tables 6 and 7 from \cite{braun2019anticipated} we perform a similar scaling to get $\mu$Jy arcmin$^{-2}$ as with MeerKAT, but we use the geometric mean of quoted minimum and maximum beam sizes. This will likely result in conservative estimates for the sensitivity.

\section{Methodology}
\label{sec:methodology}
The methodology we follow is that we generate the per-annihilation yield functions $\td{n_i}{E}{}$ for positrons, anti-protons, and photons via Monte Carlo (MC) event generators. We then use these as ingredients for our models from Section~\ref{sec:astromodel}. The model results are compared to data from cosmic-ray and gamma-ray excesses to find a best-fit parameter space, in terms of $m_\chi$ and $\langle \sigma V\rangle$. Finally, this is used to produce predictions for observable radio emissions and non-observation limits with MeerKAT and the SKA. Each of these steps is detailed below.

\subsection{Annihilation yield functions}
\begin{figure}
\begin{center}
    \includegraphics[width=0.49\hsize]{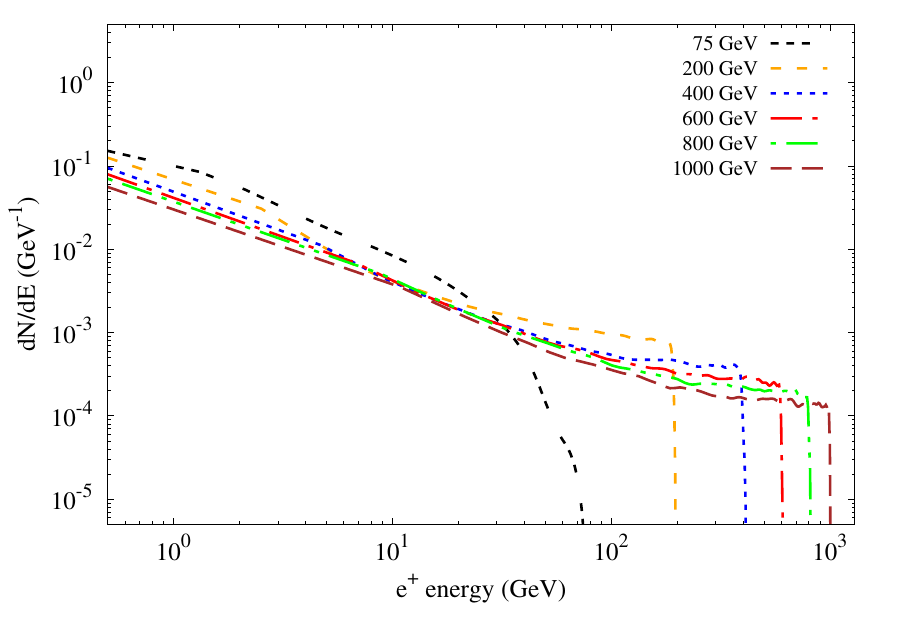} \includegraphics[width=0.49\hsize]{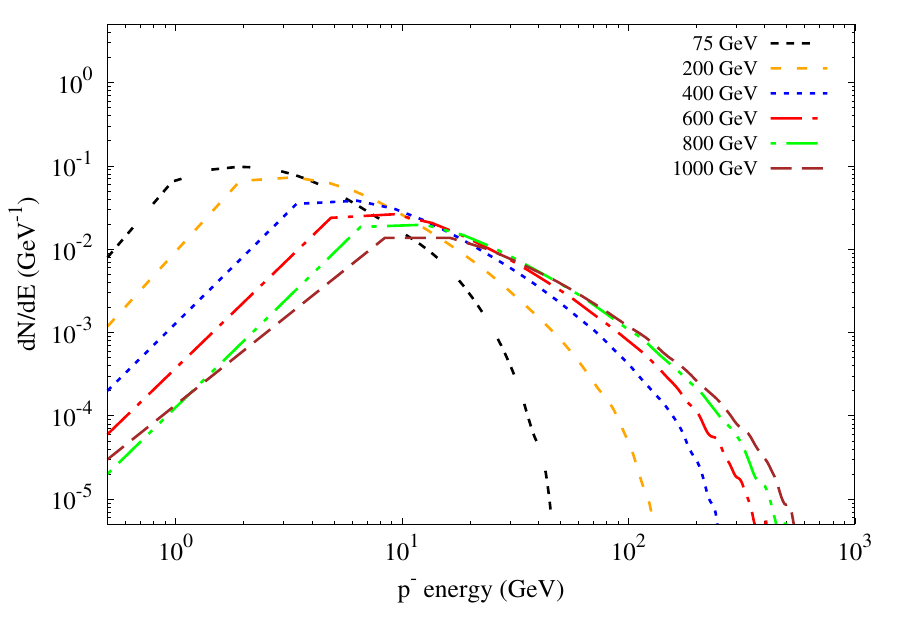}
     \includegraphics[width=0.49\hsize]{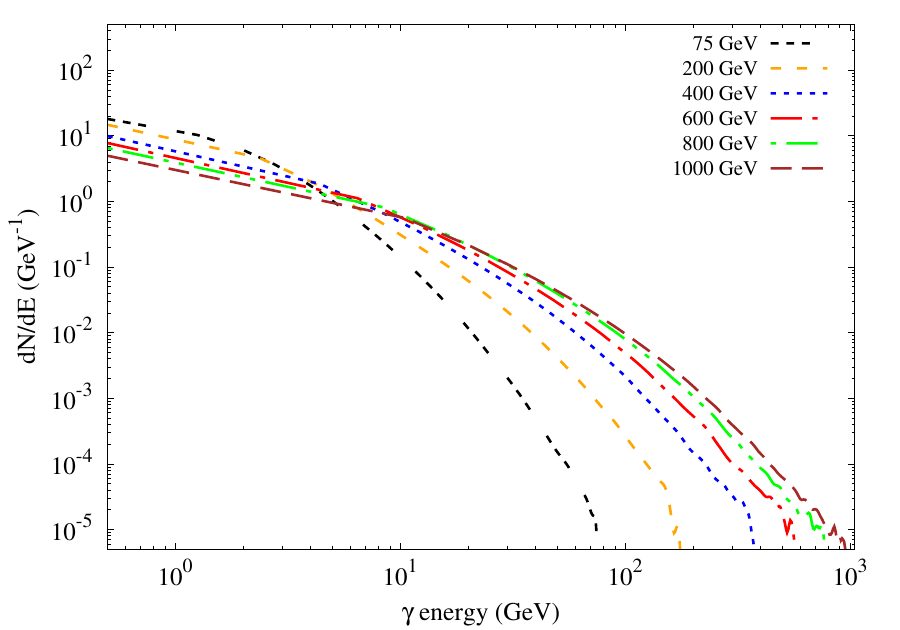}
  \caption{Differential yields of positions, anti-protons and photons for the 2$\to$2 annihilation process of Spin-0 DM for a variety of WIMP masses. Model parameters are fixed according to \cite{vonBuddenbrock:2016rmr,vonBuddenbrock:2017gvy}, based on LHC data.}
    \label{fig:yields1}
    \end{center}
\end{figure}

\begin{figure}
\begin{center}
    \includegraphics[width=0.49\hsize]{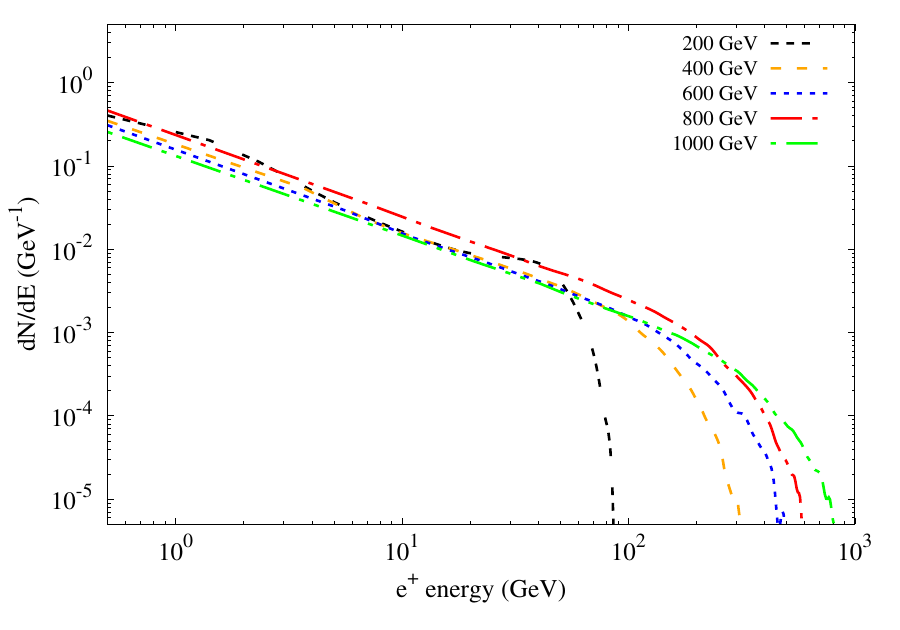} 
    \includegraphics[width=0.49\hsize]{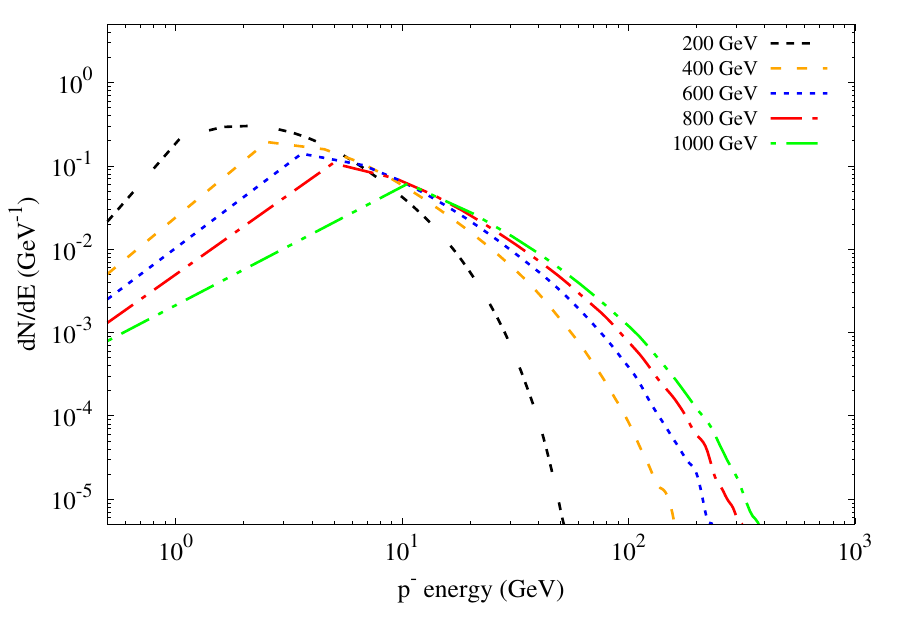}
      \includegraphics[width=0.49\hsize]{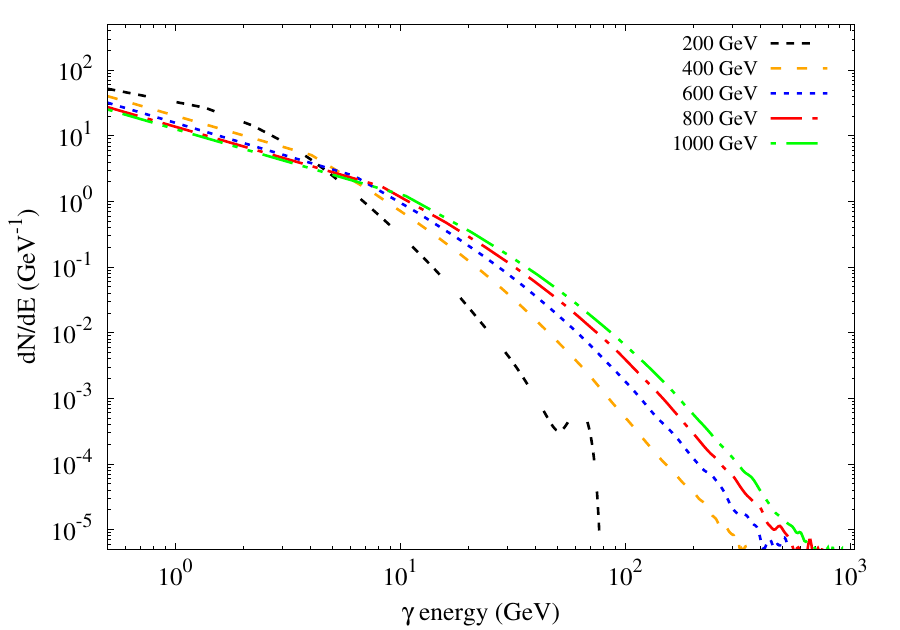}
  \caption{Differential yields of positions, anti-protons and photons for the 2$\to$3 annihilation process of Spin-0 DM for a variety of WIMP masses. Model parameters are fixed according to \cite{vonBuddenbrock:2016rmr,vonBuddenbrock:2017gvy}, based on LHC data.}
    \label{fig:yields2}
    \end{center}
\end{figure}

We use an MC generator to simulate the production of particles as a result of the annihilation of DM through $S$ according to the model described in Section~\ref{sec:model}. We make use of \texttt{MG5\_@MC} \cite{Alwall:2011uj} as our primary tool to generate events for the $2\to 2$ and $2\to 3$ scattering processes. 
The MC generator is interfaced with \texttt{Pythia 8} \cite{Sj_strand_2008} to hadronize intermediate partons. 
We varied the DM mass between 200 to 1000 GeV for the $2\to 3$ scattering and 75 to 1000 GeV for the $2\to 2$ scattering with 100 GeV spacing. In Fig.~\ref{fig:yields1} and \ref{fig:yields2} we display a complete set of yield spectra for positrons and anti-protons for 2$\to$2 and 2$\to$3 scattering. These graphs display the number of particles per annihilation per unit energy, $\td{n_i}{E}{}$, and the yield functions binned in energy intervals of 0.5 GeV. Results are displayed for spin-0 DM. A notable result is that the yield functions per annihilation are not significantly different for the other spin choices. However, the cross-section for the $2\to 3$ process is several orders of magnitude smaller than that of the $2\to 2$ case. We will thus prioritise this latter scenario in our presented results. 

\subsection{Fitting to astrophysical excesses}
The annihilation yields, found above, are then used as inputs for indirect DM detection following Eqs.~(\ref{eq:4}) and (\ref{eq:antiprotons}) in the case of the anti-proton and positron fluxes at Earth, and Eqs.~(\ref{eq:gamma}) and (\ref{eq:flux}) for gamma-rays produced via primary and secondary mechanisms. We will then use cosmic-ray and gamma-ray excess data to determine the region of the DM parameter space (in terms of $m_{\chi}$ and $\langle \sigma V\rangle$) that best fits to these excesses. The predicted spectra for cosmic-ray fluxes at Earth are compared to data from AMS-02 \cite{PhysRevLett.122.041102,Aguilar:2016kjl}. In the case of the anti-protons we use the background model found in \cite{Heisig_2020}, while for positrons we use a nearby pulsar model for the high-energy background from \cite{Fang_2019} (note that this means the DM component is sub-dominant). When we model gamma-ray fluxes from the galactic centre we use fiducial model data from \cite{Ackermann_2017}, making use of the excess spectrum for a $10^\circ$ region of interest around the galactic centre. In all cases our Milky-Way halo models follow those presented in \cite{ppdmcb1}. 

We note that both AMS-02 anomalies have been widely studied, either as constraints on DM models or as potential signatures thereof \cite{Cholis_2019,amse1,amse2,amse3,Profumo:2019pob,amse5,amsp1}. The anti-proton case has been argued to be highly significant and most in agreement with the annihilation via $b$-quarks of a $40-70$ GeV WIMP with some potential for heavier masses \cite{Cholis_2019}. However, an accounting for possibility of correlated errors \cite{Heisig_2020} reveals no preference for a DM contribution from WIMPS between 10 and 1000 GeV in mass. It is also difficult to reconcile such models with existing radio data \cite{Beck_2019}. In the case of positrons there is a stark but unexplained excess of positrons at around a few hundred GeV that can potentially be described in terms of DM models and has great potential to be used as a probe of DM physics even with astrophysical backgrounds \cite{ams2e2015}. Additionally, we note that the significance of the widely studied Fermi-LAT galactic centre gamma-ray excess is highly uncertain due to systematics \cite{Ackermann_2017}. Notably, \cite{Di_Mauro_2021} has recently argued that the galactic centre excess remains compatible with a contracted NFW \cite{nfw1996} profile (see also the review in \cite{Fischer:2021sqw}).

\subsection{Multi-frequency emission prediction}
Our final step is to use the emission models from section~\ref{sec:photons} to produce multi-frequency spectra for Reticulum II, with a focus on the radio and gamma-ray frequency bands. This we do using Eqs.~(\ref{eq:gamma}) and (\ref{eq:flux}) and by solving Eq.~(\ref{eq:difusion-eq}) following the methods detailed in section~\ref{sec:photons}. Once these predictions are produced for $\langle \sigma V\rangle$ values sourced from the excess best-fits, we compare them to the sensitivities of MeerKAT and the SKA, following the descriptions in section~\ref{sec:meerkat}, in order to determine the observability of models in the excess best-fits regions. We also determine non-observation constraints at 95\% confidence interval via the same sensitivity estimates.

\section{Results}
\label{sec:results}
We will display only results for spin 0, as $\langle \sigma V\rangle$ results here are very similar, with only a factor of two difference if the spin 1/2 case is a Dirac fermion. This is due to the yield functions per annihilation $\td{n_i}{E}{}$ being similar for all spin cases considered. The difference between the different spins occurs largely in the cross-section $\sigma$ itself. Thus, the choice of spin will largely impact limits on the coupling between DM and $S$, $g_\chi^S$. Note that we will prioritise the presentation of the $2\to 2$ scenario, due to its larger cross-section and better description of the astrophysical data.

\begin{figure}
    \centering
    \includegraphics[width=0.49\textwidth]{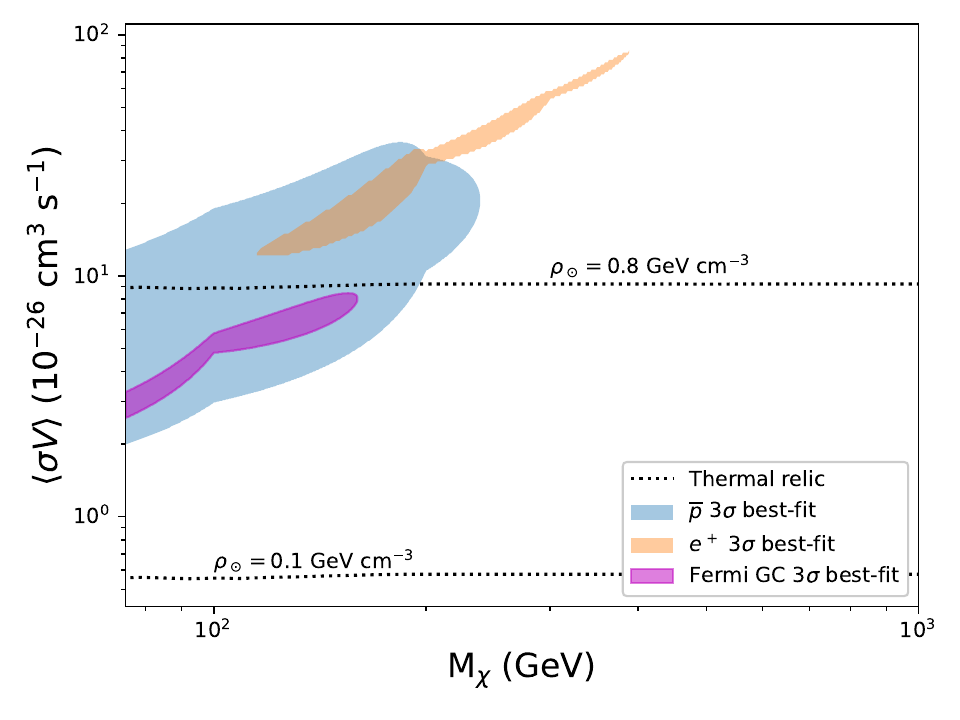}
    \includegraphics[width=0.49\hsize]{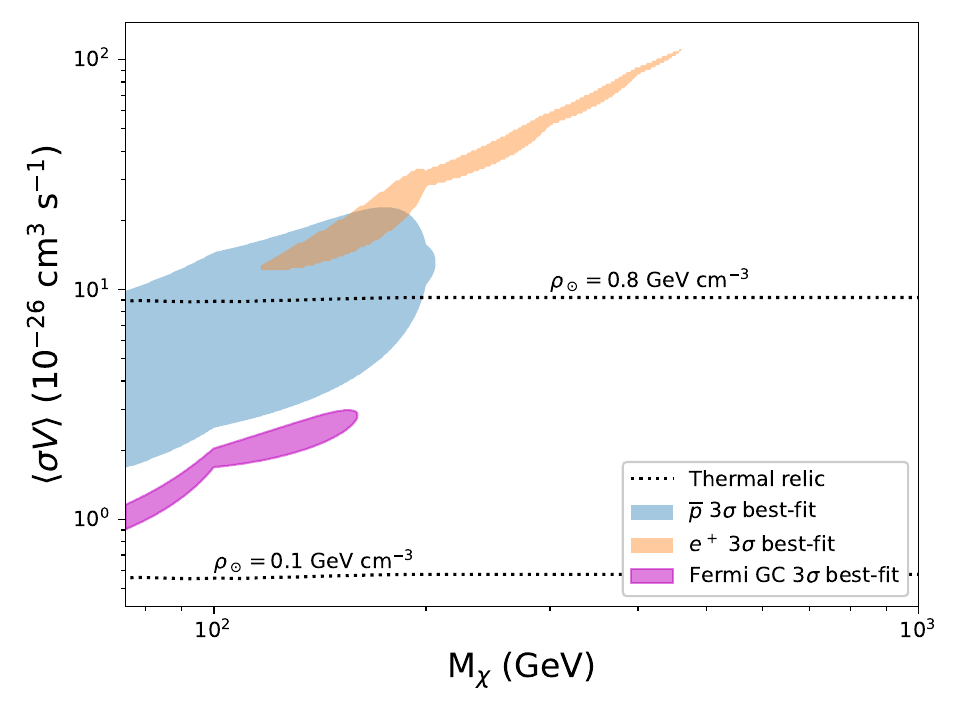}
    \includegraphics[width=0.49\hsize]{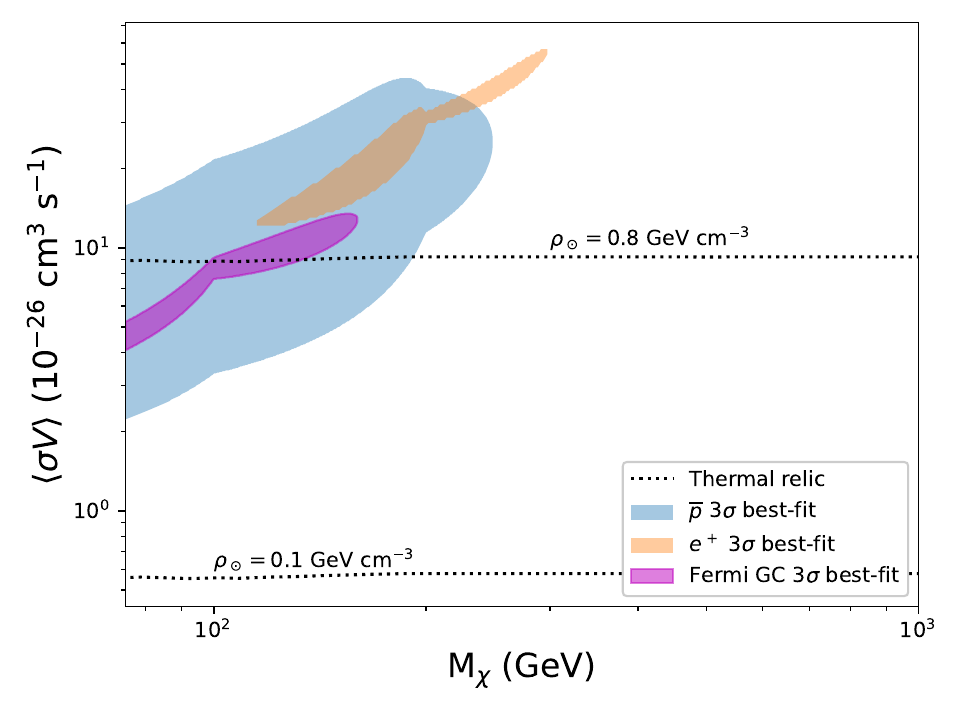}
    \caption{The parameter space fitting the $2\to2$ scattering to the Fermi-LAT galactic centre excess, AMS-02 anti-proton and positron spectra. The shaded regions represent the $3\sigma$ confidence interval. The region between the thermal relic lines represents the uncertainties in local DM density and galactic halo profile. This plot assumes the MED diffusion scenario. The 3 plots correspond to different Milky-Way DM halo profiles from \cite{ppdmcb1}. \textit{Top left}: Einasto profile. \textit{Top right}: contracted Einasto. \textit{Bottom}: NFW.}
    \label{fig:ams2Params_Ein}
\end{figure}

In Fig.~\ref{fig:ams2Params_Ein} we display the best-fit parameter space for $2\to2$ scattering with the AMS-02 positron data from \cite{PhysRevLett.122.041102} and the over-lap regions with the anti-proton and Fermi-LAT galactic centre gamma-ray parameter spaces. In order to account for uncertainties in the modelling of Milky-way DM halo we display the thermal relic cross-section \cite{steigman2012} as a band, in order to represent the fact that models within this band are compatible with the relic value up to systematic uncertainties. The uncertainties are derived entirely from estimates of $\rho_\odot$ (see for instance \cite{Weber_2010,dmlocal2019}). The plots in Fig.~\ref{fig:ams2Params_Ein} show that crucial overlap regions fall slightly above the relic band. The regions for all three excesses do not mutually overlap, although it must be noted that systematic issues with the positron and gamma-ray data \cite{PhysRevLett.122.041102,Ackermann_2017} indicate that there may be missing ingredients in the astrophysical modelling. Despite this, the 3 regions cluster close together and converge close to the relic region. The presented figure shows the best-case overlap which requires the MED diffusion scenario and an Einasto (or NFW) halo with the parameters drawn from \cite{ppdmcb1}. Note that updated diffusion parameters would increase the DM flux and thus reduce the required cross-sections slightly~\cite{G_nolini_2021}. The fact that the uncontracted Einasto and NFW halos produced a closer agreement means that this overlap does not favour a more speculative contracted halo (as is often necessary for DM modelling of these excesses historically) and is compatible with non-extremal diffusion conditions. This stands in contrast with model independent searches like \cite{Di_Mauro_2021}, where it is argued that a contracted DM profile is favoured for a DM-related explanation of the galactic centre excess spectrum from \cite{Ackermann_2017}. This difference may be down to the fact that we only make a spectral comparison to the excess, whereas cases like \cite{Di_Mauro_2021} motivate the halo shape via spatial matching as well. It is notable that, despite our inclusion of a large positron background from a nearby pulsar \cite{Fang_2019}, the favoured cross-sections remain large, indicating that DM is still playing a role in reconciling the AMS-02 positron data. We have made some initial verification that the parameter space of the 2HDM+$S$ model that describes LHC and astrophysics data are not excluded by direct DM searches. In particular, for a Type-II 2HDM+$S$ limits become weak when the ratio of the vacuum expectation values of the complex doublets, $\tan{\beta}<1$ \cite{Bell:2016ekl}, which is preferred by the LHC data \cite{vonBuddenbrock:2018xar}. This evasion of direct detection is possible as $S$ acts as a mediator between DM and the SM and is not produced directly from the SM at the LHC (it results from $H$ decays). This allows the direct signal to remain small without affecting the collider or indirect results. However, more in-depth analysis may be needed on this point. The best-fit DM parameter spaces are summarised in Tables~\ref{tab:ein_params} and \ref{tab:nfw_params}.

\begin{table} 
    \centering
    \begin{tabular}{|c|c|c|}
    \hline
    M$_\chi$ & Lower  & Upper\\
    \hline
    75.0 & 2.6 & 3.3 \\
    81.1 & 2.9 & 3.7 \\
    87.1 & 3.3 & 4.2 \\
    93.2 & 3.8 & 4.8 \\
    99.2 & 4.7 & 5.7 \\
    105.3 & 4.9 & 6.1 \\
    111.3 & 5.0 & 6.4 \\
    117.4 & 5.1 & 6.7 \\
    123.5 & 5.3 & 7.1 \\
    129.5 & 5.5 & 7.4 \\
    135.6 & 5.7 & 7.7 \\
    141.6 & 6.1 & 8.0 \\
    147.7 & 6.4 & 8.3 \\
    153.8 & 6.9 & 8.4 \\
    159.8 & 7.8 & 8.2 \\
    \hline
    \end{tabular}
    \begin{tabular}{|c|c|c|}
    \hline
    M$_\chi$ & Lower  & Upper\\
    \hline
    115.2 & 12.1 & 12.4 \\
    134.7 & 12.7 & 16.6 \\
    154.2 & 14.4 & 21.3 \\
    173.7 & 17.1 & 27.2 \\
    193.2 & 23.5 & 33.5 \\
    212.8 & 30.0 & 35.3 \\
    232.3 & 32.7 & 40.2 \\
    251.8 & 36.3 & 44.7 \\
    271.3 & 41.2 & 50.7 \\
    290.9 & 48.2 & 56.7 \\
    310.4 & 56.2 & 61.0 \\
    329.9 & 61.0 & 66.9 \\
    349.4 & 66.5 & 72.9 \\
    369.0 & 73.4 & 79.3 \\
    388.5 & 84.3 & 86.3 \\
    \hline
    \end{tabular}
    \begin{tabular}{|c|c|c|}
    \hline
    M$_\chi$ & Lower  & Upper\\
    \hline
    75.0 & 2.0 & 12.8 \\
    86.7 & 2.4 & 15.3 \\
    98.4 & 2.9 & 18.5 \\
    110.1 & 3.2 & 20.7 \\
    121.8 & 3.5 & 22.8 \\
    133.5 & 3.8 & 25.2 \\
    145.2 & 4.2 & 28.0 \\
    156.9 & 4.8 & 31.0 \\
    168.6 & 5.5 & 33.8 \\
    180.3 & 6.6 & 35.7 \\
    192.0 & 8.4 & 34.9 \\
    203.7 & 10.9 & 30.9 \\
    215.4 & 12.2 & 29.5 \\
    227.1 & 14.3 & 27.2 \\
    238.8 & 19.7 & 21.2 \\
    \hline
    \end{tabular}
    \caption{The 3$\sigma$ confidence interval best-fit parameter spaces for the 2HDM+$S$ model as well as the MED scenario, $2\to2$ processes, and an Einasto halo in the Milky-Way. The limits are presented as the upper and lower edges of contour ($10^{-26}$ cm$^3$ s$^{-1}$) for each mass (GeV). \textit{Left}: Fermi-LAT GeV gamma-rays. \textit{Centre}: AMS-02 positrons. \textit{Right}: AMS-02 anti-protons.}
    \label{tab:ein_params}
\end{table}

\begin{table} 
    \centering
    \begin{tabular}{|c|c|c|}
    \hline
    M$_\chi$ & Lower  & Upper\\
    \hline
    75.0 & 4.1 & 5.2 \\
    81.1 & 4.6 & 5.9 \\
    87.1 & 5.2 & 6.7 \\
    93.2 & 6.1 & 7.7 \\
    99.2 & 7.4 & 9.0 \\
    105.3 & 7.7 & 9.6 \\
    111.3 & 7.9 & 10.2 \\
    117.4 & 8.2 & 10.7 \\
    123.5 & 8.4 & 11.2 \\
    129.5 & 8.8 & 11.8 \\
    135.6 & 9.2 & 12.3 \\
    141.6 & 9.6 & 12.8 \\
    147.7 & 10.2 & 13.2 \\
    153.8 & 11.0 & 13.5 \\
    159.8 & 12.5 & 13.1 \\
    \hline
    \end{tabular}
    \begin{tabular}{|c|c|c|}
    \hline
    M$_\chi$ & Lower  & Upper\\
    \hline
    115.2 & 12.1 & 12.7 \\
    128.2 & 12.4 & 15.4 \\
    141.2 & 13.3 & 17.9 \\
    154.2 & 14.4 & 21.3 \\
    167.2 & 16.3 & 25.2 \\
    180.2 & 18.9 & 29.9 \\
    193.2 & 23.8 & 33.5 \\
    206.3 & 29.9 & 34.3 \\
    219.3 & 31.6 & 36.7 \\
    232.3 & 33.5 & 40.2 \\
    245.3 & 35.9 & 43.2 \\
    258.3 & 38.5 & 47.0 \\
    271.3 & 42.2 & 50.7 \\
    284.4 & 46.9 & 54.4 \\
    297.4 & 54.4 & 57.0 \\
    \hline
    \end{tabular}
    \begin{tabular}{|c|c|c|}
    \hline
    M$_\chi$ & Lower  & Upper\\
    \hline
    75.0 & 2.2 & 14.5 \\
    87.4 & 2.7 & 17.5 \\
    99.8 & 3.3 & 21.6 \\
    112.2 & 3.6 & 24.0 \\
    124.7 & 4.0 & 26.9 \\
    137.1 & 4.4 & 30.2 \\
    149.5 & 5.0 & 33.9 \\
    161.9 & 5.7 & 38.1 \\
    174.3 & 6.7 & 42.1 \\
    186.7 & 8.3 & 44.2 \\
    199.1 & 11.1 & 41.0 \\
    211.5 & 12.6 & 39.4 \\
    224.0 & 14.5 & 37.6 \\
    236.4 & 17.1 & 34.4 \\
    248.8 & 24.4 & 26.1 \\
    \hline
    \end{tabular}
    \caption{The 3$\sigma$ confidence interval best-fit parameter spaces for the 2HDM+$S$ model as well as the MED scenario, $2\to2$ processes, and an NFW halo in the Milky-Way. The limits are presented as the upper and lower edges of contour ($10^{-26}$ cm$^3$ s$^{-1}$) for each mass (GeV). \textit{Left}: Fermi-LAT GeV gamma-rays. \textit{Centre}: AMS-02 positrons. \textit{Right}: AMS-02 anti-protons.}
    \label{tab:nfw_params}
\end{table}

Next, we will consider the comparison between the sensitivity of the MeerKAT telescope and the predicted emissions for our constrained DM model using the Reticulum II dwarf galaxy. The choice of this target is motivated by the fact that it is in an ideal location for southern hemisphere observations and has one of the highest `J-factors' among known dwarf galaxies \cite{Albert_2017}. In fitting with our AMS-02 constraints from Fig.~\ref{fig:ams2Params_Ein}, we assume $\langle \sigma V\rangle = 10^{-25}$~cm$^3$~s$^{-1}$. Note that we also include a shaded uncertainty band in our plots, this includes uncertainties on the value of $\rho_\odot$, the magnetic field, and the J-factor. We make use of a cored density profile following arguments from \cite{walker2009,adams2014}. This being an Einasto profile \cite{einasto1968} (which can be cored for certain parameter choices), given by:
\begin{align}
   \rho_{e}(r)=\rho_{s} \exp\left[-\frac{2}{\alpha} \left(\left[\frac{r}{r_s}\right]^{\alpha} - 1\right)\right],
\end{align}
where we follow \cite{regis2017} in having $\alpha = 0.4$, $r_s = 0.2$ kpc, and $\rho_s = 7\times 10^7$ M$_\odot$ kpc$^{-3}$. 
We then follow \cite{regis2017} in using the profiles for gas density and magnetic field strength:
\begin{equation}
n_e (r) = n_0 \exp\left(-\frac{r}{r_d}\right),
\, B (r) = B_0 \exp\left(-\frac{r}{r_d}\right).
\end{equation}
We take $r_d$ to be given by the stellar-half-light radius with a value of $35$ pc \cite{bechtol2015,koposov2015} and we assume $B_0 \approx 1$~$\mu$G, $n_0 \approx 10^{-6}$~cm$^{-3}$. The region of flux integration are taken as 3 arcminutes for radio and 30 arcminutes for gamma-rays. The first is in keeping with our assumption of arcminute scale emissions when determining sensitivity. The second is to reflect Fermi-LAT observations from \cite{Albert_2017}.

\begin{figure} 
    \centering
    \includegraphics[width=0.49\hsize]{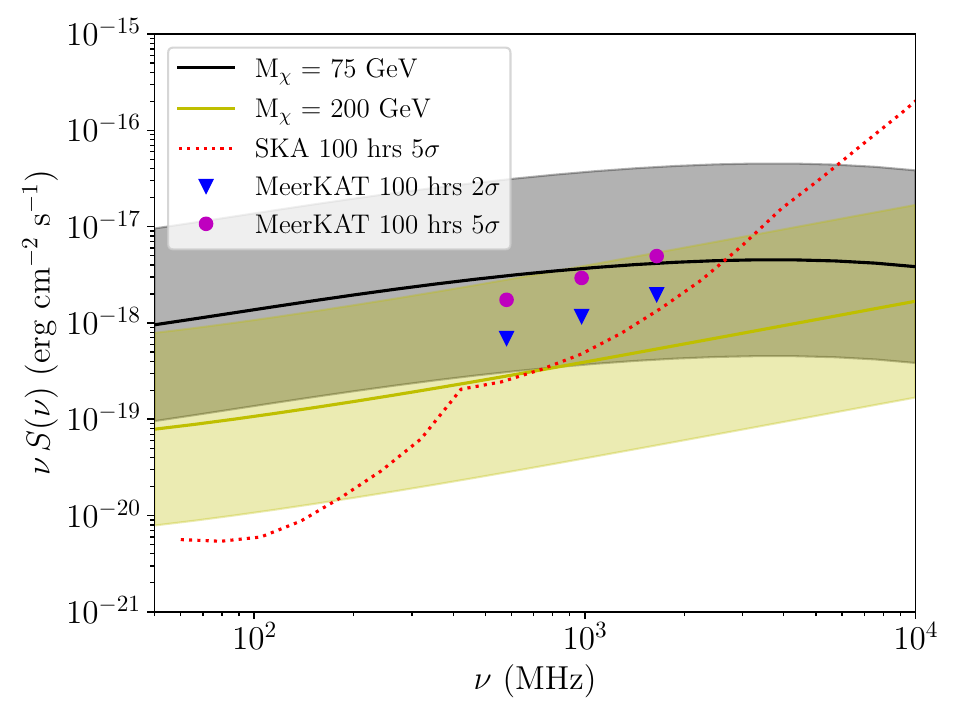}
    \includegraphics[width=0.49\hsize]{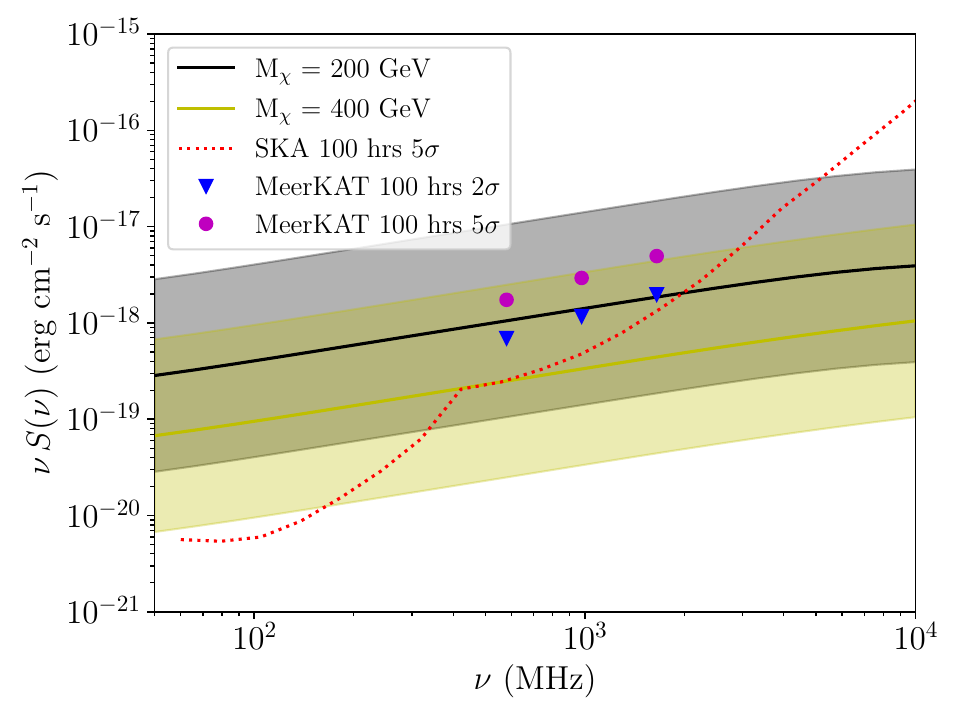}
    \caption{Radio spectrum prediction for Reticulum II with the Einasto profile. The shaded regions encompass the cross-section uncertainties from \ref{fig:ams2Params_Ein}, as well as those from the J-factor of the halo and magnetic field. \textit{Left}: $2\to2$ scattering. \textit{Right}: $2\to3$ scattering.}
    \label{fig:meerKATSpectrumEin}
\end{figure}

Fig.~\ref{fig:meerKATSpectrumEin} shows that Reticulum II with the Einasto profile produces radio fluxes detectable at $2\sigma$ confidence interval with MeerKAT across the mass range (although the $2\to2$ 200 GeV case is very marginal). Only the lower masses are detectable at $5\sigma$. However, the entire uncertainty band is not covered in either case. It should noted that the $2\to3$ case presents slightly weaker emissions (and is not 5$\sigma$ detectable). Importantly, we also display the estimated sensitivity of the full SKA \cite{braun2019anticipated} which manages to probe the entire uncertainty region for both $2\to2$ and $2\to3$ processes at $5\sigma$ confidence interval within 100 hours of observing time. 

\begin{figure} 
    \centering
    \includegraphics[width=0.49\hsize]{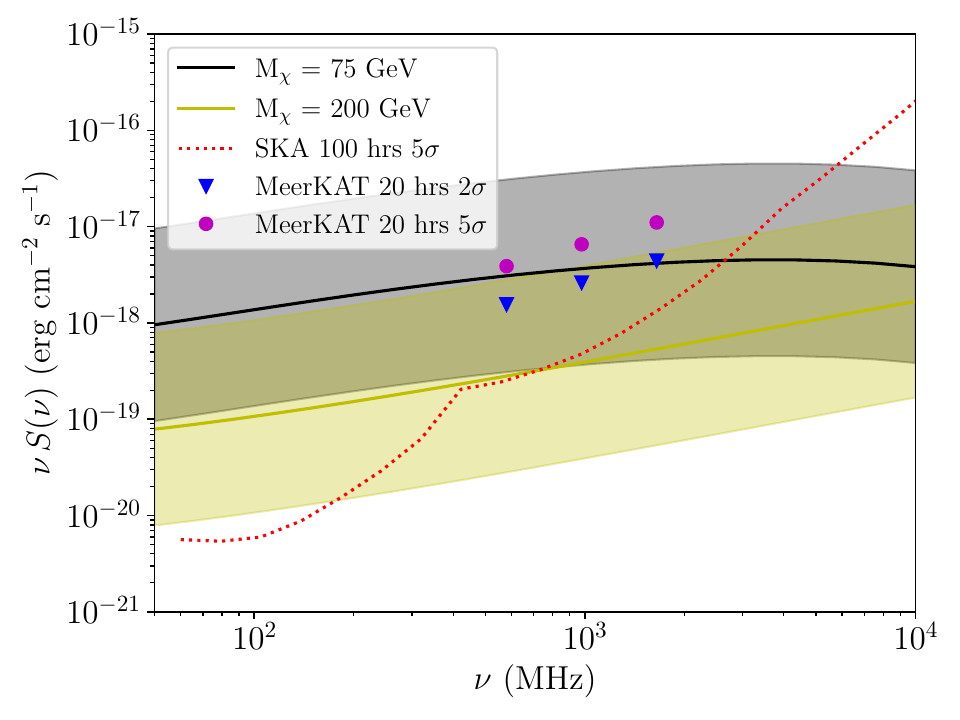}
    \includegraphics[width=0.49\hsize]{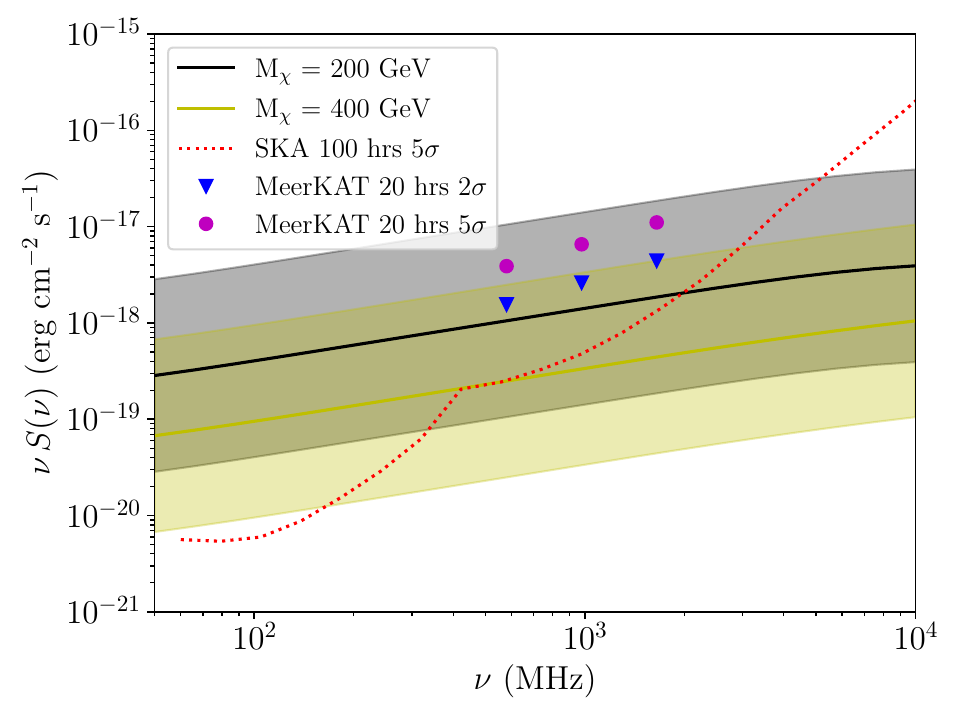}
    \caption{Radio spectrum prediction for Reticulum II with the Einasto profile. The shaded regions encompass the cross-section uncertainties from \ref{fig:ams2Params_Ein}, as well as those from the J-factor of the halo and magnetic field. \textit{Left}: $2\to2$ scattering. \textit{Right}: $2\to3$ scattering.}
    \label{fig:meerKATSpectrumEin2}
\end{figure}

In Fig.~\ref{fig:meerKATSpectrumEin2} we consider the case of a more realistic observing time of 20 hours with MeerKAT. The most notable result here is that the solid-line model within the uncertainty band ($\rho_\odot = 0.4$ GeV cm$^{-3}$, $B_0 = 1$ $\mu$G, $J \approx 2\times 10^{19}$ GeV$^2$ cm$^{-5}$) is still MeerKAT-detectable at 2 $\sigma$ for $m_\chi = 75$ GeV with substantial uncertainty band coverage at 2$\sigma$. The 5$\sigma$ level is somewhat marginal. The $2\to 3$ case is only observable for the upper portion of the uncertainty band. This should demonstrate that observations of targets like Reticulum II with MeerKAT would prove to be fruitful first steps in probing the 2HDM+$S$ model.  

For any MeerKAT $5\sigma$ detection prospects across the mass range (i.e. $m_\chi \gtrsim 200$ GeV), at least $1600$ hours of observation time would be necessary. Notably this when $B_0 = 1$ $\mu$G and $m_\chi = 200$ GeV. The fact that detection would require such vast lengths of MeerKAT observation time suggests that this is improbable when $B_0 = 1$ $\mu$G. However, an increase in sensitivity to $0.5$ $\mu$Jy beam$^{-1}$ would drop the required time to at least 40 hours. Thus, obtaining sub-micro-Jansky sensitivities starts to bring about the practical possibility of detection. This, however, would require the full SKA \cite{braun2019anticipated}. It should be noted that sufficient sensitivity for $2\sigma$ exclusion of 2HDM+$S$ model at the fiducial relic cross-section, with the lowest studied DM mass, can be obtained with at least 100 hours on MeerKAT or 6 hours on the sub-micro-Jansky case. The time required to probe a 200 GeV WIMP at the relic level is around one hundred times longer than the 75 GeV case. This is largely due to MeerKAT's limited frequency coverage, which will be ameliorated in the full SKA. 

One point to note is that of the search prospects will likely improve when a visibility taper is considered, as this suppresses the long array baselines, tuning the sensitivity towards larger scale emissions \cite{radioAstronFund}. This consideration makes the presented results somewhat conservative. Another important issue is the magnetic field. At present there is no data on dwarf galaxy magnetic environments. However, theoretical arguments suggest that the magnetic field should be at least $B\sim 0.4$ $\mu$G~\cite{Regis_2014} and that the Milky-Way magnetic field has a magnitude of $\approx 1.4$ $\mu$G at the location of Reticulum II~\cite{regis2017}. This variation has been taken into account in the shaded regions.

\begin{figure} 
    \centering
    \includegraphics[width=0.49\hsize]{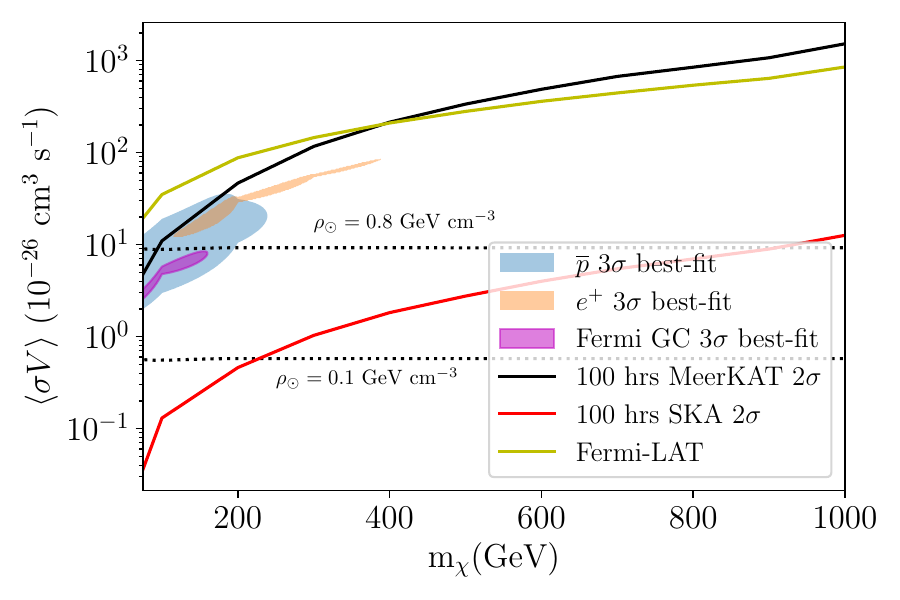}
    \includegraphics[width=0.49\hsize]{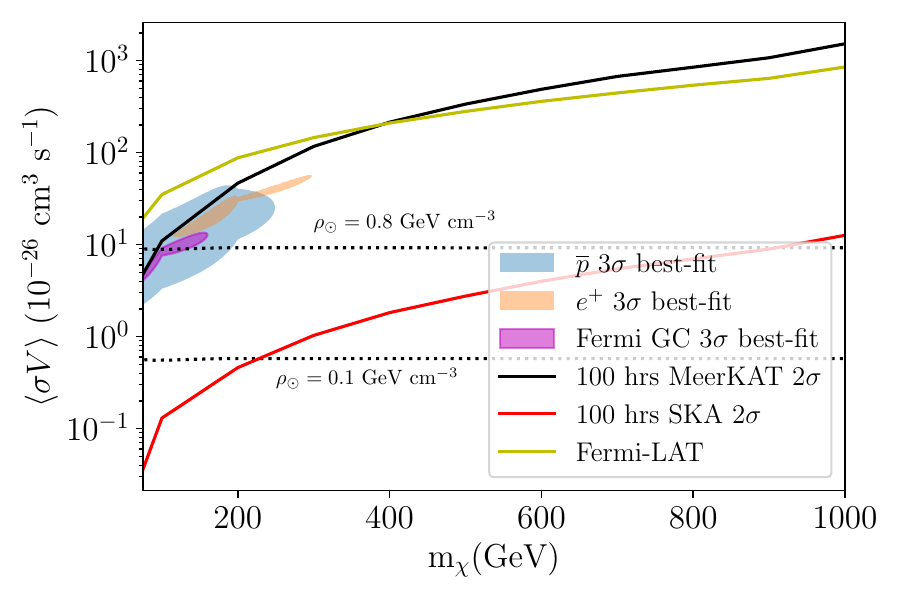}
    \caption{Non-observation $2\sigma$ exclusion projections for Reticulum II (Einasto profile) with MeerKAT and the SKA. \textit{Left}: Milkyway-Way galactic centre with Einasto profile. \textit{Right}: Milkyway-Way galactic centre with NFW profile. The shaded regions are drawn from Fig.~\ref{fig:ams2Params_Ein}. The Fermi-LAT limits are drawn from data found in \cite{Albert_2017} and we use a $30^\prime$ integration region in this case.}
    \label{fig:meerKATParams}
\end{figure}

In Fig.~\ref{fig:meerKATParams} we display the potential non-observation constraints from 100 hours of observation of Reticulum II with both MeerKAT and the SKA. MeerKAT cannot probe the 2HDM+$S$ model down to the thermal relic cross-section for all DM masses considered. However, it can exclude a part of the excess overlap regions for the astrophysical excesses when $\rho_\odot = 0.4$ GeV cm$^{-3}$, in both cases of NFW and Einasto density profiles for the Milky-Way. The current uncertainties in the local DM density mean the overlap region from Fig.~\ref{fig:ams2Params_Ein} can potentially evade MeerKAT exclusion with 100 hours of observation time if the local density is $\rho_\odot \sim 0.8$ GeV cm$^{-3}$. This is, however, an extreme scenario. Thus, MeerKAT shows considerable potential to probe the parameter space corresponding to the 2HDM+$S$ DM candidate accounting for the studied astrophysical excesses. The potential upper limits are summarised in table~\ref{tab:limits}. Thus, the two-fold sensitivity advantage over previous observations of Reticulum II suggest that even 20 hours on target would yield cutting edge model-independent constraints on channels like $b$ quarks, as \cite{regis2017} obtained highly competitive non-observation results with 30 hours of ATCA time on this same galaxy.

\begin{table} 
    \centering
    \begin{tabular}{|c|c|c|}
    \hline
    M$_\chi$ &  MeerKAT &  SKA \\
    (GeV)& ($10^{-26}$ cm$^3$ s$^{-1}$) & ($10^{-26}$ cm$^3$ s$^{-1}$)\\
    \hline
    75  &  $4.8$ &   $0.036$ \\
    100 &   $11$  &  $0.13$ \\
    200 &   $47$  &  $0.46$ \\
    300 &   $1.2\times 10^2$  &  $1.0$ \\
    400 &   $2.1\times 10^2$  &  $1.8$ \\
    500 &   $3.4\times 10^2$  &  $2.8$ \\
    600 &   $4.9\times 10^2$  &  $4.0$ \\
    700 &   $6.7\times 10^2$  &  $5.5$ \\
    800 &   $8.5\times 10^2$  &  $7.0$ \\
    900 &   $1.1\times 10^3$  &  $9.0$ \\
    1000 &  $1.5\times 10^3$  &  $13$ \\
    \hline
    \end{tabular}
    \caption{Potential non-observation upper limits on $\langle \sigma V\rangle$ for the 2HDM+$S$ model in the $2\to2$ scenario. This data assumes the MED diffusion scenario and an Einasto halo profile in the Milky-Way.}
    \label{tab:limits}
\end{table}

\begin{figure}
    \centering
    \includegraphics[width=0.49\hsize]{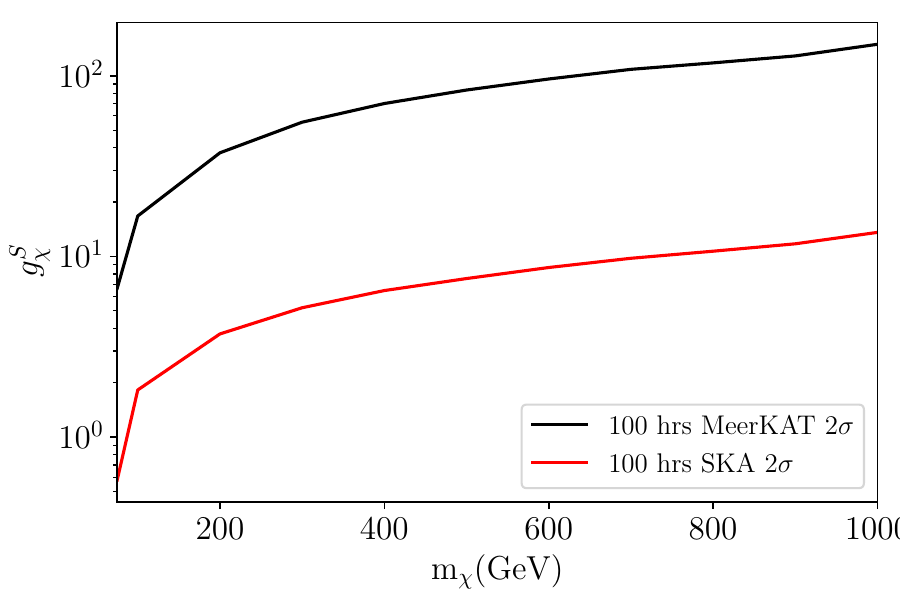}
    \includegraphics[width=0.49\hsize]{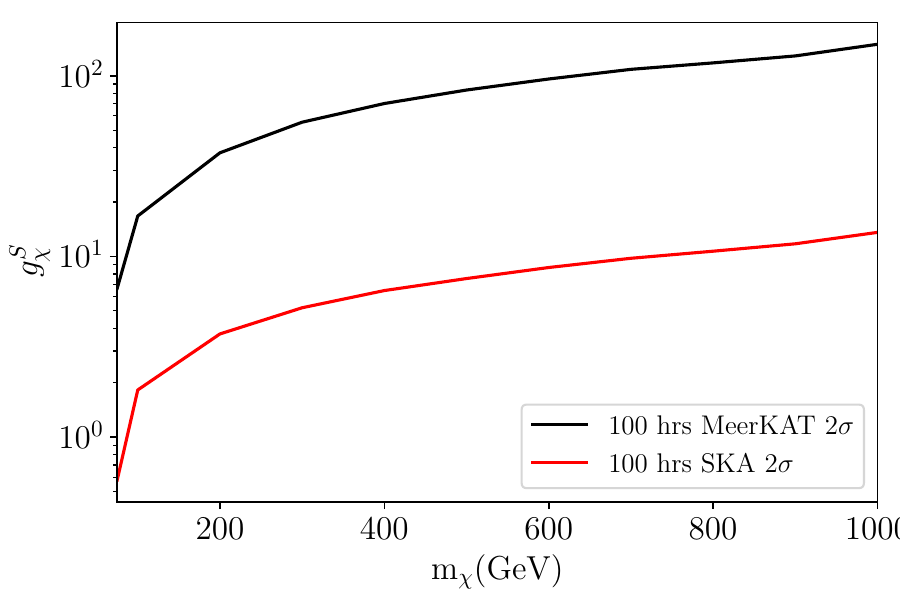}
    \caption{Non-observation $2\sigma$ exclusion projections on the coupling between $S$ and $\chi$ ($2\to 2$ case) for Reticulum II (Einasto profile) with MeerKAT and the SKA. \textit{Left}: Milky-Way galactic centre with Einasto profile. \textit{Right}: Milky-Way galactic centre with NFW profile.}
    \label{fig:meerKATCoupling}
\end{figure}

Next we will translate limits on $\langle \sigma V\rangle$ from fig.~\ref{fig:meerKATParams} to the coupling between $S$ and $\chi$, $g_\chi^S$, this we will do via the definition
\begin{equation}
    \langle \sigma V\rangle = \int \sigma_{\chi\chi}(g_\chi^S=1) \left(g^S_\chi\right)^2 v f(v) d v \; , 
\end{equation}
where $\sigma_{\chi\chi}(g_\chi^S=1)$ is the annihilation cross-section when the coupling is unity and $f(v)$ is a Maxwell-Boltzmann distribution with velocity dispersion $4$ km s$^{-1}$ to reflect the Reticulum II halo \cite{koposov2015}.
In figure~\ref{fig:meerKATCoupling} we display the limits on the coupling parameter between $\chi$ and $S$ that can be inferred from the results of figure~\ref{fig:meerKATParams} in the $2\to 2$ case with spin 0. These demonstrate that this parameter space remains difficult to probe, however, for low $m_{\chi}$ we see that the SKA has the potential to probe below $g^S_\chi = 1$ (the $2\to 3$ case provides no real limits as the cross-section is around $10^6$ times smaller, suggesting it would be subdominant anyway). It is notable that this coupling is unconstrained by current LHC data. Importantly, spin 1/2 yields similar results whereas spin 1 results are less constraining by several orders of magnitude at every mass considered. This is due to the far smaller $\chi\chi \to S$ cross-section for spin 1 $\chi$.

A final important question is how generic are these results? Can they be applied to other WIMP scenarios? In fact, this is not possible, as the annihilation yields displayed in Figs.~\ref{fig:yields1} and \ref{fig:yields2} do not correspond strongly to any of the generic channels ($b\bar{b}$, $W^+W^-$, $l^+l^-$, $hh$ etc). The mapping from 2HDM+$S$ yields to these channels depends on both the mass of the WIMP and energy of the product particle. This makes translating these results to more general chases challenging. 

\section{Conclusions}
\label{sec:conclusions}
In this paper we used a 2HDM+$S$ model, that describes the multi-lepton anomalies at the LHC, to describe the excesses in gamma-ray flux from the galactic centre and the cosmic-ray spectra from AMS-02. This is achieved through DM annihilation via the singlet scalar into particles of the SM. The parameters of the model are fixed to the LHC data, except for the mass of the DM and the size of the coupling to the mediator $S$ (these being, as yet, unconstrained by collider data). The mass of the DM is scanned, where the coupling of the DM to the mediator $S$ is varied, and various diffusion scenarios are considered. A satisfactory description of the gamma-ray flux from the galactic centre and the cosmic-ray spectra from AMS-02 is obtained with the MED diffusion scenario. The best description of the excesses is obtained for DM masses below 200 GeV. Although complete overlap between all the best-fit regions is not achieved, it is notable that there are unresolved systematics in much of the data. Nonetheless, it is still remarkable that such close agreement can be obtained for a model whose particle physics is set by LHC anomalies. 

Predictions of the synchrotron spectrum are made with the model in order to assess the detection sensitivity of MeerKAT in a conservative scenario where no visibility taper is considered. We conducted our test on a fiducial best-fit cross-section of $10^{-25}$ cm$^3$ s$^{-1}$ to characterise the region of overlap for the various excesses, and made use of only a 3 arcminute observation region and scaled point source sensitivity for MeerKAT. We found that $5\sigma$ detection is marginally possible, within 20 hours of observation at lower masses. However, around 1000 hours might be necessary to survey up to 200 GeV. This would still fail to cover the entire uncertainty region, so some of the parameter space can evade any conceivable detection with MeerKAT. This is not the case with the full SKA. Which, at $5\sigma$ confidence interval and 100 hours observing time, can probe the entire uncertainty band for all masses in the excess overlap window. However, we note that an optimised MeerKAT search for diffuse emissions will likely have increased sensitivity.

It is worth highlighting that the MeerKAT sensitivity estimates have highly competitive constraining power, even somewhat exceeding that of Fermi-LAT data. MeerKAT can explore a significant area of the full excess overlap parameter space at $2\sigma$ confidence level with 100 hours of observing time. Whereas, Fermi-LAT with data from \cite{Albert_2017} and a 10 times larger integration radius does not intrude at all upon these parameter regions. On the other hand, the full SKA can explore the entire overlap parameter space, even within uncertainties due to $\rho_\odot$, with less than 100 hours of observation. However, at present MeerKAT will be the frontier in radio instruments for DM searches, especially since it has begun observing calls already and will be upgraded in the near future with $~20\%$ more dishes. 

The results documented in this work have implications on searches for DM at the LHC, due to the fact that limits on the cross-section from $\chi\chi \to$ SM translate directly to the coupling between $\chi$ and $S$ in Eq.~(\ref{intdm}), as the other model parameters are fixed. Additionally, as an SM singlet, $S$ is predominantly produced via the decay of the heavy scalar $H\rightarrow SS$. Thus, DM produced via the decay of $S$ can recoil against SM particles that $S$ can also decay into \cite{vonBuddenbrock:2016rmr}. Of particular interest would be the resonant search for $S\rightarrow ZZ,Z\gamma,\gamma\gamma$ in association with moderate missing transverse energy carried by the DM. The astrophysical data provides a consistency test as well as a narrowed region of focus, in which collider data can be used to further probe the properties of 2HDM+$S$ with respect to its unconstrained degrees of freedom. 

\section*{Acknowledgments}
The authors want to thank Andreas Crivellin and Bhupal Dev for most useful discussions. The authors are grateful for support from the South African Department of Science and Innovation through the SA-CERN program and the National Research Foundation for various forms of support. GB acknowledges the funding of the National Research Foundation through Thuthuka grant number 117969.

\section*{Data availability}
All the data used to produce this work is available from the authors on reasonable request.

\bibliographystyle{elsarticle-num}

\bibliography{apssamp}

\providecommand{\noopsort}[1]{}\providecommand{\singleletter}[1]{#1}%
\begin{thebibliography}{10}
\expandafter\ifx\csname url\endcsname\relax
  \def\url#1{\texttt{#1}}\fi
\expandafter\ifx\csname urlprefix\endcsname\relax\def\urlprefix{URL }\fi
\expandafter\ifx\csname href\endcsname\relax
  \def\href#1#2{#2} \def\path#1{#1}\fi

\bibitem{Higgs:1964ia}
P.~W. Higgs, {Broken symmetries, massless particles and gauge fields}, Phys.
  Lett. 12 (1964) 132--133.
\newblock \href {https://doi.org/10.1016/0031-9163(64)91136-9}
  {\path{doi:10.1016/0031-9163(64)91136-9}}.

\bibitem{Englert:1964et}
F.~Englert, R.~Brout, {Broken Symmetry and the Mass of Gauge Vector Mesons},
  Phys. Rev. Lett. 13 (1964) 321--323.
\newblock \href {https://doi.org/10.1103/PhysRevLett.13.321}
  {\path{doi:10.1103/PhysRevLett.13.321}}.

\bibitem{Higgs:1964pj}
P.~W. Higgs, {Broken Symmetries and the Masses of Gauge Bosons}, Phys. Rev.
  Lett. 13 (1964) 508--509.
\newblock \href {https://doi.org/10.1103/PhysRevLett.13.508}
  {\path{doi:10.1103/PhysRevLett.13.508}}.

\bibitem{Guralnik:1964eu}
G.~S. Guralnik, C.~R. Hagen, T.~W.~B. Kibble, {Global Conservation Laws and
  Massless Particles}, Phys. Rev. Lett. 13 (1964) 585--587.
\newblock \href {https://doi.org/10.1103/PhysRevLett.13.585}
  {\path{doi:10.1103/PhysRevLett.13.585}}.

\bibitem{Aad:2012tfa}
G.~Aad, et~al., {Observation of a new particle in the search for the Standard
  Model Higgs boson with the ATLAS detector at the LHC}, Phys. Lett. B 716
  (2012) 1--29.
\newblock \href {http://arxiv.org/abs/1207.7214} {\path{arXiv:1207.7214}},
  \href {https://doi.org/10.1016/j.physletb.2012.08.020}
  {\path{doi:10.1016/j.physletb.2012.08.020}}.

\bibitem{Chatrchyan:2012ufa}
S.~Chatrchyan, et~al., {Observation of a New Boson at a Mass of 125 GeV with
  the CMS Experiment at the LHC}, Phys. Lett. B 716 (2012) 30--61.
\newblock \href {http://arxiv.org/abs/1207.7235} {\path{arXiv:1207.7235}},
  \href {https://doi.org/10.1016/j.physletb.2012.08.021}
  {\path{doi:10.1016/j.physletb.2012.08.021}}.

\bibitem{Chatrchyan:2012jja}
S.~Chatrchyan, et~al., {Study of the Mass and Spin-Parity of the Higgs Boson
  Candidate Via Its Decays to Z Boson Pairs}, Phys. Rev. Lett. 110~(8) (2013)
  081803.
\newblock \href {http://arxiv.org/abs/1212.6639} {\path{arXiv:1212.6639}},
  \href {https://doi.org/10.1103/PhysRevLett.110.081803}
  {\path{doi:10.1103/PhysRevLett.110.081803}}.

\bibitem{Aad:2013xqa}
G.~Aad, et~al., {Evidence for the spin-0 nature of the Higgs boson using ATLAS
  data}, Phys. Lett. B 726 (2013) 120--144.
\newblock \href {http://arxiv.org/abs/1307.1432} {\path{arXiv:1307.1432}},
  \href {https://doi.org/10.1016/j.physletb.2013.08.026}
  {\path{doi:10.1016/j.physletb.2013.08.026}}.

\bibitem{vonBuddenbrock:2016rmr}
S.~von Buddenbrock, N.~Chakrabarty, A.~S. Cornell, D.~Kar, M.~Kumar, T.~Mandal,
  B.~Mellado, B.~Mukhopadhyaya, R.~G. Reed, X.~Ruan, {Phenomenological
  signatures of additional scalar bosons at the LHC}, Eur. Phys. J. C 76~(10)
  (2016) 580.
\newblock \href {http://arxiv.org/abs/1606.01674} {\path{arXiv:1606.01674}},
  \href {https://doi.org/10.1140/epjc/s10052-016-4435-8}
  {\path{doi:10.1140/epjc/s10052-016-4435-8}}.

\bibitem{vonBuddenbrock:2017gvy}
S.~von Buddenbrock, A.~S. Cornell, A.~Fadol, M.~Kumar, B.~Mellado, X.~Ruan,
  {Multi-lepton signatures of additional scalar bosons beyond the Standard
  Model at the LHC}, J. Phys. G 45~(11) (2018) 115003.
\newblock \href {http://arxiv.org/abs/1711.07874} {\path{arXiv:1711.07874}},
  \href {https://doi.org/10.1088/1361-6471/aae3d6}
  {\path{doi:10.1088/1361-6471/aae3d6}}.

\bibitem{vonBuddenbrock:2019ajh}
S.~Buddenbrock, A.~S. Cornell, Y.~Fang, A.~Fadol~Mohammed, M.~Kumar,
  B.~Mellado, K.~G. Tomiwa, {The emergence of multi-lepton anomalies at the LHC
  and their compatibility with new physics at the EW scale}, JHEP 10 (2019)
  157.
\newblock \href {http://arxiv.org/abs/1901.05300} {\path{arXiv:1901.05300}},
  \href {https://doi.org/10.1007/JHEP10(2019)157}
  {\path{doi:10.1007/JHEP10(2019)157}}.

\bibitem{vonBuddenbrock:2020ter}
S.~von Buddenbrock, R.~Ruiz, B.~Mellado, {Anatomy of inclusive $t\bar t W$
  production at hadron colliders}, Phys. Lett. B 811 (2020) 135964.
\newblock \href {http://arxiv.org/abs/2009.00032} {\path{arXiv:2009.00032}},
  \href {https://doi.org/10.1016/j.physletb.2020.135964}
  {\path{doi:10.1016/j.physletb.2020.135964}}.

\bibitem{Hernandez:2019geu}
Y.~Hernandez, M.~Kumar, A.~S. Cornell, S.-E. Dahbi, Y.~Fang, B.~Lieberman,
  B.~Mellado, K.~Monnakgotla, X.~Ruan, S.~Xin, {The anomalous production of
  multi-lepton and its impact on the measurement of $Wh$ production at the
  LHC}, Eur. Phys. J. C 81~(4) (2021) 365.
\newblock \href {http://arxiv.org/abs/1912.00699} {\path{arXiv:1912.00699}},
  \href {https://doi.org/10.1140/epjc/s10052-021-09137-1}
  {\path{doi:10.1140/epjc/s10052-021-09137-1}}.

\bibitem{Buddenbrock:2019tua}
S.~Buddenbrock, A.~S. Cornell, Y.~Fang, A.~Fadol~Mohammed, M.~Kumar,
  B.~Mellado, K.~G. Tomiwa, {The emergence of multi-lepton anomalies at the LHC
  and their compatibility with new physics at the EW scale}, JHEP 10 (2019)
  157.
\newblock \href {http://arxiv.org/abs/1901.05300} {\path{arXiv:1901.05300}},
  \href {https://doi.org/10.1007/JHEP10(2019)157}
  {\path{doi:10.1007/JHEP10(2019)157}}.

\bibitem{Crivellin:2021ubm}
A.~Crivellin, Y.~Fang, O.~Fischer, A.~Kumar, M.~Kumar, E.~Malwa, B.~Mellado,
  N.~Rapheeha, X.~Ruan, Q.~Sha, {Accumulating Evidence for the Associate
  Production of a Neutral Scalar with Mass around 151 GeV} (9 2021).
\newblock \href {http://arxiv.org/abs/2109.02650} {\path{arXiv:2109.02650}}.

\bibitem{Fischer:2021sqw}
O.~Fischer, B.~Mellado, S.~Antusch, E.~Bagnaschi, S.~Banerjee, G.~Beck,
  B.~Belfatto, M.~Bellis, Z.~Berezhiani, M.~Blanke, B.~Capdevila, K.~Cheung,
  A.~Crivellin, N.~Desai, B.~Dev, R.~Godbole, T.~Han, P.~Harris,
  M.~Hoferichter, M.~Kirk, S.~Kulkarni, C.~Lange, K.~Lassila-Perini, Z.~Liu,
  F.~Mahmoudi, C.~A. Manzari, D.~Marzocca, B.~Mukhopadhyaya, A.~Pich, X.~Ruan,
  L.~Schnell, J.~Thaler, S.~Westhoff,
  \href{https://doi.org/10.1140%2Fepjc%2Fs10052-022-10541-4}{Unveiling hidden
  physics at the {LHC}}, The European Physical Journal C 82~(8) (aug 2022).
\newblock \href {https://doi.org/10.1140/epjc/s10052-022-10541-4}
  {\path{doi:10.1140/epjc/s10052-022-10541-4}}.
\newline\urlprefix\url{https://doi.org/10.1140%2Fepjc%2Fs10052-022-10541-4}

\bibitem{Sabatta:2019nfg}
D.~Sabatta, A.~S. Cornell, A.~Goyal, M.~Kumar, B.~Mellado, X.~Ruan, {Connecting
  muon anomalous magnetic moment and multi-lepton anomalies at LHC}, Chin.
  Phys. C 44~(6) (2020) 063103.
\newblock \href {http://arxiv.org/abs/1909.03969} {\path{arXiv:1909.03969}},
  \href {https://doi.org/10.1088/1674-1137/44/6/063103}
  {\path{doi:10.1088/1674-1137/44/6/063103}}.

\bibitem{Jueid:2021avn}
A.~Jueid, J.~Kim, S.~Lee, J.~Song, {Type-X two-Higgs-doublet model in light of
  the muon g-2: Confronting Higgs boson and collider data}, Phys. Rev. D
  104~(9) (2021) 095008.
\newblock \href {http://arxiv.org/abs/2104.10175} {\path{arXiv:2104.10175}},
  \href {https://doi.org/10.1103/PhysRevD.104.095008}
  {\path{doi:10.1103/PhysRevD.104.095008}}.

\bibitem{Kanemura:2021dez}
S.~Kanemura, M.~Takeuchi, K.~Yagyu, {Probing double-aligned two-Higgs-doublet
  models at the LHC}, Phys. Rev. D 105~(11) (2022) 115001.
\newblock \href {http://arxiv.org/abs/2112.13679} {\path{arXiv:2112.13679}},
  \href {https://doi.org/10.1103/PhysRevD.105.115001}
  {\path{doi:10.1103/PhysRevD.105.115001}}.

\bibitem{Fischer:2021nha}
O.~Fischer, M.~Lindner, S.~van~der Woude, {Robustness of ARS leptogenesis in
  scalar extensions}, JHEP 05 (2022) 149.
\newblock \href {http://arxiv.org/abs/2110.14499} {\path{arXiv:2110.14499}},
  \href {https://doi.org/10.1007/JHEP05(2022)149}
  {\path{doi:10.1007/JHEP05(2022)149}}.

\bibitem{Antusch:2021oit}
S.~Antusch, O.~Fischer, A.~Hammad, C.~Scherb, {Explaining excesses in
  four-leptons at the LHC with a double peak from a CP violating Two Higgs
  Doublet Model}, JHEP 08 (2022) 224.
\newblock \href {http://arxiv.org/abs/2112.00921} {\path{arXiv:2112.00921}},
  \href {https://doi.org/10.1007/JHEP08(2022)224}
  {\path{doi:10.1007/JHEP08(2022)224}}.

\bibitem{Aboubrahim:2021phn}
A.~Aboubrahim, P.~Nath, R.~M. Syed, {Yukawa coupling unification in an SO(10)
  model consistent with Fermilab (g \ensuremath{-} 2)$_{\mu}$ result}, JHEP 06
  (2021) 002.
\newblock \href {http://arxiv.org/abs/2104.10114} {\path{arXiv:2104.10114}},
  \href {https://doi.org/10.1007/JHEP06(2021)002}
  {\path{doi:10.1007/JHEP06(2021)002}}.

\bibitem{2020Planck}
N.~Aghanim, Y.~Akrami, M.~Ashdown, J.~Aumont, C.~Baccigalupi, M.~Ballardini,
  A.~J. Banday, R.~B. Barreiro, N.~Bartolo, et~al.,
  \href{http://dx.doi.org/10.1051/0004-6361/201833910}{Planck 2018 results},
  Astronomy \& Astrophysics 641 (2020) A6.
\newblock \href {https://doi.org/10.1051/0004-6361/201833910}
  {\path{doi:10.1051/0004-6361/201833910}}.
\newline\urlprefix\url{http://dx.doi.org/10.1051/0004-6361/201833910}

\bibitem{Koopmans_2003}
L.~V.~E. Koopmans, T.~Treu, \href{https://dx.doi.org/10.1086/345423}{The
  structure and dynamics of luminous and dark matter in the early‐type lens
  galaxy of 0047−281 at z= 0.485}, The Astrophysical Journal 583~(2) (2003)
  606–615.
\newblock \href {https://doi.org/10.1086/345423} {\path{doi:10.1086/345423}}.
\newline\urlprefix\url{https://dx.doi.org/10.1086/345423}

\bibitem{Metcalf_2004}
R.~B. Metcalf, L.~A. Moustakas, A.~J. Bunker, I.~R. Parry,
  \href{http://dx.doi.org/10.1086/383243}{Spectroscopic gravitational lensing
  and limits on the dark matter substructure in q2237+0305}, The Astrophysical
  Journal 607~(1) (2004) 43–59.
\newblock \href {https://doi.org/10.1086/383243} {\path{doi:10.1086/383243}}.
\newline\urlprefix\url{http://dx.doi.org/10.1086/383243}

\bibitem{Hoekstra_2002}
H.~Hoekstra, H.~Yee, M.~D. Gladders,
  \href{http://dx.doi.org/10.1016/S1387-6473(02)00245-2}{Current status of weak
  gravitational lensing}, New Astronomy Reviews 46~(12) (2002) 767–781.
\newblock \href {https://doi.org/10.1016/s1387-6473(02)00245-2}
  {\path{doi:10.1016/s1387-6473(02)00245-2}}.
\newline\urlprefix\url{http://dx.doi.org/10.1016/S1387-6473(02)00245-2}

\bibitem{Moustakas_2003}
L.~A. Moustakas, R.~B. Metcalf,
  \href{http://dx.doi.org/10.1046/j.1365-8711.2003.06055.x}{Detecting dark
  matter substructure spectroscopically in strong gravitational lenses},
  Monthly Notices of the Royal Astronomical Society 339~(3) (2003) 607–615.
\newblock \href {https://doi.org/10.1046/j.1365-8711.2003.06055.x}
  {\path{doi:10.1046/j.1365-8711.2003.06055.x}}.
\newline\urlprefix\url{http://dx.doi.org/10.1046/j.1365-8711.2003.06055.x}

\bibitem{vonBuddenbrock:2018xar}
S.~von Buddenbrock, A.~S. Cornell, E.~D.~R. Iarilala, M.~Kumar, B.~Mellado,
  X.~Ruan, E.~M. Shrif, {Constraints on a 2HDM with a singlet scalar and
  implications in the search for heavy bosons at the LHC}, J. Phys. G 46~(11)
  (2019) 115001.
\newblock \href {http://arxiv.org/abs/1809.06344} {\path{arXiv:1809.06344}},
  \href {https://doi.org/10.1088/1361-6471/ab3cf6}
  {\path{doi:10.1088/1361-6471/ab3cf6}}.

\bibitem{PhysRevLett.122.041102}
M.~Aguilar, et~al.,
  \href{https://link.aps.org/doi/10.1103/PhysRevLett.122.041102}{Towards
  understanding the origin of cosmic-ray positrons}, Phys. Rev. Lett. 122
  (2019) 041102.
\newblock \href {https://doi.org/10.1103/PhysRevLett.122.041102}
  {\path{doi:10.1103/PhysRevLett.122.041102}}.
\newline\urlprefix\url{https://link.aps.org/doi/10.1103/PhysRevLett.122.041102}

\bibitem{Aguilar:2016kjl}
M.~Aguilar, et~al., {Antiproton Flux, Antiproton-to-Proton Flux Ratio, and
  Properties of Elementary Particle Fluxes in Primary Cosmic Rays Measured with
  the Alpha Magnetic Spectrometer on the International Space Station}, Phys.
  Rev. Lett. 117~(9) (2016) 091103.
\newblock \href {https://doi.org/10.1103/PhysRevLett.117.091103}
  {\path{doi:10.1103/PhysRevLett.117.091103}}.

\bibitem{Ackermann_2017}
M.~Ackermann, M.~Ajello, A.~Albert, W.~B. Atwood, L.~Baldini, J.~Ballet,
  G.~Barbiellini, D.~Bastieri, R.~Bellazzini, E.~Bissaldi, et~al.,
  \href{http://dx.doi.org/10.3847/1538-4357/aa6cab}{The fermi galactic center
  gev excess and implications for dark matter}, The Astrophysical Journal
  840~(1) (2017) 43.
\newblock \href {https://doi.org/10.3847/1538-4357/aa6cab}
  {\path{doi:10.3847/1538-4357/aa6cab}}.
\newline\urlprefix\url{http://dx.doi.org/10.3847/1538-4357/aa6cab}

\bibitem{Cholis_2019}
I.~Cholis, T.~Linden, D.~Hooper,
  \href{http://dx.doi.org/10.1103/PhysRevD.99.103026}{A robust excess in the
  cosmic-ray antiproton spectrum: Implications for annihilating dark matter},
  Physical Review D 99~(10) (May 2019).
\newblock \href {https://doi.org/10.1103/physrevd.99.103026}
  {\path{doi:10.1103/physrevd.99.103026}}.
\newline\urlprefix\url{http://dx.doi.org/10.1103/PhysRevD.99.103026}

\bibitem{amse1}
A.~Das, B.~Dasgupta, A.~Ray,
  \href{https://link.aps.org/doi/10.1103/PhysRevD.101.063014}{Galactic positron
  excess from selectively enhanced dark matter annihilation}, Phys. Rev. D 101
  (2020) 063014.
\newblock \href {https://doi.org/10.1103/PhysRevD.101.063014}
  {\path{doi:10.1103/PhysRevD.101.063014}}.
\newline\urlprefix\url{https://link.aps.org/doi/10.1103/PhysRevD.101.063014}

\bibitem{amse2}
K.~Ishiwata, O.~Macias, S.~Ando, M.~Arimoto,
  \href{https://doi.org/10.1088%2F1475-7516%2F2020%2F01%2F003}{Probing heavy
  dark matter decays with multi-messenger astrophysical data}, Journal of
  Cosmology and Astroparticle Physics 2020~(01) (2020) 003--003.
\newblock \href {https://doi.org/10.1088/1475-7516/2020/01/003}
  {\path{doi:10.1088/1475-7516/2020/01/003}}.
\newline\urlprefix\url{https://doi.org/10.1088%2F1475-7516%2F2020%2F01%2F003}

\bibitem{amse3}
Y.~Farzan, M.~Rajaee,
  \href{https://doi.org/10.1088%2F1475-7516%2F2019%2F04%2F040}{Dark matter
  decaying into millicharged particles as a solution to {AMS}-02 positron
  excess}, Journal of Cosmology and Astroparticle Physics 2019~(04) (2019)
  040--040.
\newblock \href {https://doi.org/10.1088/1475-7516/2019/04/040}
  {\path{doi:10.1088/1475-7516/2019/04/040}}.
\newline\urlprefix\url{https://doi.org/10.1088%2F1475-7516%2F2019%2F04%2F040}

\bibitem{Profumo:2019pob}
S.~Profumo, F.~Queiroz, C.~Siqueira, {Has AMS-02 Observed Two-Component Dark
  Matter?}, J. Phys. G 48~(1) (2020) 015006.
\newblock \href {http://arxiv.org/abs/1903.07638} {\path{arXiv:1903.07638}},
  \href {https://doi.org/10.1088/1361-6471/abbd20}
  {\path{doi:10.1088/1361-6471/abbd20}}.

\bibitem{amse5}
T.~Li, \href{http://dx.doi.org/10.1007/JHEP01(2018)151}{Revisiting simplified
  dark matter models in terms of ams-02 and fermi-lat}, Journal of High Energy
  Physics 2018~(1) (Jan 2018).
\newblock \href {https://doi.org/10.1007/jhep01(2018)151}
  {\path{doi:10.1007/jhep01(2018)151}}.
\newline\urlprefix\url{http://dx.doi.org/10.1007/JHEP01(2018)151}

\bibitem{amsp1}
G.~Giesen, M.~Boudaud, Y.~Génolini, V.~Poulin, M.~Cirelli, P.~Salati, P.~D.
  Serpico, \href{http://dx.doi.org/10.1088/1475-7516/2015/09/023}{Ams-02
  antiprotons, at last! secondary astrophysical component and immediate
  implications for dark matter}, Journal of Cosmology and Astroparticle Physics
  2015~(09) (2015) 023–023.
\newblock \href {https://doi.org/10.1088/1475-7516/2015/09/023}
  {\path{doi:10.1088/1475-7516/2015/09/023}}.
\newline\urlprefix\url{http://dx.doi.org/10.1088/1475-7516/2015/09/023}

\bibitem{Beck_2019}
G.~Beck, \href{http://dx.doi.org/10.1088/1475-7516/2019/08/019}{An excess of
  excesses examined via dark matter radio emissions from galaxies}, Journal of
  Cosmology and Astroparticle Physics 2019~(08) (2019) 019–019.
\newblock \href {https://doi.org/10.1088/1475-7516/2019/08/019}
  {\path{doi:10.1088/1475-7516/2019/08/019}}.
\newline\urlprefix\url{http://dx.doi.org/10.1088/1475-7516/2019/08/019}

\bibitem{gs2016}
G.~Beck, S.~Colafrancesco, {A Multi-frequency analysis of dark matter
  annihilation interpretations of recent anti-particle and $\gamma$-ray
  excesses in cosmic structures}, JCAP 1605~(05) (2016) 013.
\newblock \href {http://arxiv.org/abs/1508.01386} {\path{arXiv:1508.01386}},
  \href {https://doi.org/10.1088/1475-7516/2016/05/013}
  {\path{doi:10.1088/1475-7516/2016/05/013}}.

\bibitem{CMS:2014suk}
S.~Chatrchyan, et~al., {Evidence for the direct decay of the 125 GeV Higgs
  boson to fermions}, Nature Phys. 10 (2014) 557--560.
\newblock \href {http://arxiv.org/abs/1401.6527} {\path{arXiv:1401.6527}},
  \href {https://doi.org/10.1038/nphys3005} {\path{doi:10.1038/nphys3005}}.

\bibitem{CMS:2015hra}
V.~Khachatryan, et~al., {Search for a Higgs boson in the mass range from 145 to
  1000 GeV decaying to a pair of W or Z bosons}, JHEP 10 (2015) 144.
\newblock \href {http://arxiv.org/abs/1504.00936} {\path{arXiv:1504.00936}},
  \href {https://doi.org/10.1007/JHEP10(2015)144}
  {\path{doi:10.1007/JHEP10(2015)144}}.

\bibitem{CMS:2015mba}
{Search for H/A decaying into Z+A/H, with Z to ll and A/H to fermion pair}
  (2015).

\bibitem{ppdmcb1}
M.~Cirelli, et~al., Pppc 4 dm id: A poor particle physicist cookbook for dark
  matter indirect detection, JCAP 1103 (2011) 051.

\bibitem{maurin2001cosmic}
D.~Maurin, F.~Donato, R.~Taillet, P.~Salati, Cosmic rays below z= 30 in a
  diffusion model: new constraints on propagation parameters, The Astrophysical
  Journal 555~(2) (2001) 585.

\bibitem{G_nolini_2021}
Y.~G{\'{e} }nolini, M.~Boudaud, M.~Cirelli, L.~Derome, J.~Lavalle, D.~Maurin,
  P.~Salati, N.~Weinrich,
  \href{https://doi.org/10.1103%2Fphysrevd.104.083005}{New minimal, median, and
  maximal propagation models for dark matter searches with galactic cosmic
  rays}, Physical Review D 104~(8) (oct 2021).
\newblock \href {https://doi.org/10.1103/physrevd.104.083005}
  {\path{doi:10.1103/physrevd.104.083005}}.
\newline\urlprefix\url{https://doi.org/10.1103%2Fphysrevd.104.083005}

\bibitem{baltz2004}
E.~A. Baltz, L.~Wai,
  \href{https://link.aps.org/doi/10.1103/PhysRevD.70.023512}{Diffuse inverse
  compton and synchrotron emission from dark matter annihilations in galactic
  satellites}, Physical Review D 70 (2004) 023512.
\newblock \href {https://doi.org/10.1103/PhysRevD.70.023512}
  {\path{doi:10.1103/PhysRevD.70.023512}}.
\newline\urlprefix\url{https://link.aps.org/doi/10.1103/PhysRevD.70.023512}

\bibitem{baltz1999}
E.~A. Baltz, J.~Edsj\"o,
  \href{https://link.aps.org/doi/10.1103/PhysRevD.59.023511}{Positron
  propagation and fluxes from neutralino annihilation in the halo}, Physical
  Review D 59 (1998) 023511.
\newblock \href {https://doi.org/10.1103/PhysRevD.59.023511}
  {\path{doi:10.1103/PhysRevD.59.023511}}.
\newline\urlprefix\url{https://link.aps.org/doi/10.1103/PhysRevD.59.023511}

\bibitem{Colafrancesco2006}
S.~Colafrancesco, S.~Profumo, P.~Ullio, Multi-frequency analysis of neutralino
  dark matter annihilations in the coma cluster, Astronomy and Astrophysics 455
  (2006) 21.

\bibitem{Colafrancesco2007}
S.~Colafrancesco, S.~Profumo, P.~Ullio, Detecting dark matter wimps in the
  draco dwarf: a multi-wavelength perspective, Physical Review D 75 (2007)
  023513.

\bibitem{Regis_2014}
M.~Regis, S.~Colafrancesco, S.~Profumo, W.~de~Blok, M.~Massardi, L.~Richter,
  \href{http://dx.doi.org/10.1088/1475-7516/2014/10/016}{Local group dsph radio
  survey with atca (iii): constraints on particle dark matter}, Journal of
  Cosmology and Astroparticle Physics 2014~(10) (2014) 016–016.
\newblock \href {https://doi.org/10.1088/1475-7516/2014/10/016}
  {\path{doi:10.1088/1475-7516/2014/10/016}}.
\newline\urlprefix\url{http://dx.doi.org/10.1088/1475-7516/2014/10/016}

\bibitem{egorov2013}
A.~E. {Egorov}, E.~{Pierpaoli}, {Constraints on dark matter annihilation by
  radio observations of M31}, Physical Review D 88~(2) (2013) 023504.
\newblock \href {http://arxiv.org/abs/1304.0517} {\path{arXiv:1304.0517}},
  \href {https://doi.org/10.1103/PhysRevD.88.023504}
  {\path{doi:10.1103/PhysRevD.88.023504}}.

\bibitem{longair1994}
M.~S. Longair, High Energy Astrophysics, Cambridge University Press, 1994.

\bibitem{rybicki1986}
G.~B. {Rybicki}, A.~P. {Lightman}, {Radiative Processes in Astrophysics},
  Wiley, 1986.

\bibitem{atwood2009large}
W.~Atwood, A.~A. Abdo, M.~Ackermann, W.~Althouse, B.~Anderson, M.~Axelsson,
  L.~Baldini, J.~Ballet, D.~Band, G.~Barbiellini, et~al., The large area
  telescope on the fermi gamma-ray space telescope mission, The Astrophysical
  Journal 697~(2) (2009) 1071.

\bibitem{ska2012}
P.~Dewdney, W.~Turner, R.~Millenaar, R.~McCool, J.~Lazio, T.~Cornwell, Ska
  baseline design document (2012).

\bibitem{gsp2015}
S.~Colafrancesco, P.~Marchegiani, G.~Beck, Evolution of dark matter halos and
  their radio emissions, JCAP 02 (2015) 032C.

\bibitem{regis2017}
M.~Regis, L.~Richter, S.~Colafrancesco,
  \href{http://stacks.iop.org/1475-7516/2017/i=07/a=025}{Dark matter in the
  reticulum ii dsph: a radio search}, Journal of Cosmology and Astroparticle
  Physics 2017~(07) (2017) 025.
\newline\urlprefix\url{http://stacks.iop.org/1475-7516/2017/i=07/a=025}

\bibitem{radioAstronFund}
J.~Marr, R.~Snell, S.~Kurtz, Fundamentals of Radio Astronomy: Observational
  Methods, CRC Press, 2015.

\bibitem{booth2009meerkat}
R.~S. Booth, W.~J.~G. de~Blok, J.~L. Jonas, B.~Fanaroff, Meerkat key project
  science, specifications, and proposals (2009).
\newblock \href {http://arxiv.org/abs/0910.2935} {\path{arXiv:0910.2935}}.

\bibitem{braun2019anticipated}
R.~Braun, A.~Bonaldi, T.~Bourke, E.~Keane, J.~Wagg, Anticipated performance of
  the square kilometre array -- phase 1 (ska1) (2019).
\newblock \href {http://arxiv.org/abs/1912.12699} {\path{arXiv:1912.12699}}.

\bibitem{Alwall:2011uj}
J.~Alwall, M.~Herquet, F.~Maltoni, O.~Mattelaer, T.~Stelzer, {MadGraph 5 :
  Going Beyond}, JHEP 06 (2011) 128.
\newblock \href {http://arxiv.org/abs/1106.0522} {\path{arXiv:1106.0522}},
  \href {https://doi.org/10.1007/JHEP06(2011)128}
  {\path{doi:10.1007/JHEP06(2011)128}}.

\bibitem{Sj_strand_2008}
T.~Sjöstrand, S.~Mrenna, P.~Skands,
  \href{http://dx.doi.org/10.1016/j.cpc.2008.01.036}{A brief introduction to
  pythia 8.1}, Computer Physics Communications 178~(11) (2008) 852–867.
\newblock \href {https://doi.org/10.1016/j.cpc.2008.01.036}
  {\path{doi:10.1016/j.cpc.2008.01.036}}.
\newline\urlprefix\url{http://dx.doi.org/10.1016/j.cpc.2008.01.036}

\bibitem{Heisig_2020}
J.~Heisig, M.~Korsmeier, M.~W. Winkler,
  \href{http://dx.doi.org/10.1103/PhysRevResearch.2.043017}{Dark matter or
  correlated errors: Systematics of the ams-02 antiproton excess}, Physical
  Review Research 2~(4) (Oct 2020).
\newblock \href {https://doi.org/10.1103/physrevresearch.2.043017}
  {\path{doi:10.1103/physrevresearch.2.043017}}.
\newline\urlprefix\url{http://dx.doi.org/10.1103/PhysRevResearch.2.043017}

\bibitem{Fang_2019}
K.~Fang, X.-J. Bi, P.-F. Yin,
  \href{http://dx.doi.org/10.3847/1538-4357/ab3fac}{Reanalysis of the pulsar
  scenario to explain the cosmic positron excess considering the recent
  developments}, The Astrophysical Journal 884~(2) (2019) 124.
\newblock \href {https://doi.org/10.3847/1538-4357/ab3fac}
  {\path{doi:10.3847/1538-4357/ab3fac}}.
\newline\urlprefix\url{http://dx.doi.org/10.3847/1538-4357/ab3fac}

\bibitem{ams2e2015}
M.~D. Mauro, F.~Donato, N.~Fornengo, A.~Vittino,
  \href{http://dx.doi.org/10.1088/1475-7516/2016/05/031}{Dark matter vs.
  astrophysics in the interpretation of ams-02 electron and positron data},
  Journal of Cosmology and Astroparticle Physics 2016~(05) (2016) 031–031.
\newblock \href {https://doi.org/10.1088/1475-7516/2016/05/031}
  {\path{doi:10.1088/1475-7516/2016/05/031}}.
\newline\urlprefix\url{http://dx.doi.org/10.1088/1475-7516/2016/05/031}

\bibitem{Di_Mauro_2021}
M.~Di~Mauro,
  \href{http://dx.doi.org/10.1103/PhysRevD.103.063029}{Characteristics of the
  galactic center excess measured with 11 years of fermi -lat data}, Physical
  Review D 103~(6) (2021) 063029.
\newblock \href {https://doi.org/10.1103/physrevd.103.063029}
  {\path{doi:10.1103/physrevd.103.063029}}.
\newline\urlprefix\url{http://dx.doi.org/10.1103/PhysRevD.103.063029}

\bibitem{nfw1996}
J.~F. Navarro, C.~S. Frenk, S.~D.~M. White, {The Structure of cold dark matter
  halos}, The Astrophysical Journal 462 (1996) 563--575.
\newblock \href {http://arxiv.org/abs/astro-ph/9508025}
  {\path{arXiv:astro-ph/9508025}}, \href {https://doi.org/10.1086/177173}
  {\path{doi:10.1086/177173}}.

\bibitem{steigman2012}
G.~Steigman, B.~Dasgupta, J.~F. Beacom, {Precise Relic WIMP Abundance and its
  Impact on Searches for Dark Matter Annihilation}, Phys. Rev. D86 (2012)
  023506.
\newblock \href {http://arxiv.org/abs/1204.3622} {\path{arXiv:1204.3622}},
  \href {https://doi.org/10.1103/PhysRevD.86.023506}
  {\path{doi:10.1103/PhysRevD.86.023506}}.

\bibitem{Weber_2010}
M.~Weber, W.~de~Boer,
  \href{http://dx.doi.org/10.1051/0004-6361/200913381}{Determination of the
  local dark matter density in our galaxy}, Astronomy and Astrophysics 509
  (2010) A25.
\newblock \href {https://doi.org/10.1051/0004-6361/200913381}
  {\path{doi:10.1051/0004-6361/200913381}}.
\newline\urlprefix\url{http://dx.doi.org/10.1051/0004-6361/200913381}

\bibitem{dmlocal2019}
M.~Benito, A.~Cuoco, F.~Iocco,
  \href{http://dx.doi.org/10.1088/1475-7516/2019/03/033}{Handling the
  uncertainties in the galactic dark matter distribution for particle dark
  matter searches}, Journal of Cosmology and Astroparticle Physics 2019~(03)
  (2019) 033–033.
\newblock \href {https://doi.org/10.1088/1475-7516/2019/03/033}
  {\path{doi:10.1088/1475-7516/2019/03/033}}.
\newline\urlprefix\url{http://dx.doi.org/10.1088/1475-7516/2019/03/033}

\bibitem{Bell:2016ekl}
N.~F. Bell, G.~Busoni, I.~W. Sanderson, {Self-consistent Dark Matter Simplified
  Models with an s-channel scalar mediator}, JCAP 03 (2017) 015.
\newblock \href {http://arxiv.org/abs/1612.03475} {\path{arXiv:1612.03475}},
  \href {https://doi.org/10.1088/1475-7516/2017/03/015}
  {\path{doi:10.1088/1475-7516/2017/03/015}}.

\bibitem{Albert_2017}
A.~Albert, et~al.,
  \href{http://dx.doi.org/10.3847/1538-4357/834/2/110}{Searching for dark
  matter annihilation in recently discovered milky way satellites with
  fermi-lat}, The Astrophysical Journal 834~(2) (2017) 110.
\newblock \href {https://doi.org/10.3847/1538-4357/834/2/110}
  {\path{doi:10.3847/1538-4357/834/2/110}}.
\newline\urlprefix\url{http://dx.doi.org/10.3847/1538-4357/834/2/110}

\bibitem{walker2009}
M.~G. Walker, M.~Mateo, E.~W. Olszewski, J.~P. narrubia, N.~W. Evans,
  G.~Gilmore, \href{http://stacks.iop.org/0004-637X/704/i=2/a=1274}{A universal
  mass profile for dwarf spheroidal galaxies?}, ApJ 704~(2) (2009) 1274.
\newline\urlprefix\url{http://stacks.iop.org/0004-637X/704/i=2/a=1274}

\bibitem{adams2014}
J.~J. Adams, et~al., \href{http://stacks.iop.org/0004-637X/789/i=1/a=63}{Dwarf
  galaxy dark matter density profiles inferred from stellar and gas
  kinematics}, ApJ 789~(1) (2014) 63.
\newline\urlprefix\url{http://stacks.iop.org/0004-637X/789/i=1/a=63}

\bibitem{einasto1968}
J.~Einasto, On galactic descriptive functions, Publications of the Tartuskoj
  Astrofizica Observatory 36 (1968) 414.

\bibitem{bechtol2015}
K.~{Bechtol}, et~al., {Eight New Milky Way Companions Discovered in First-year
  Dark Energy Survey Data}, ApJ 807 (2015) 50.
\newblock \href {http://arxiv.org/abs/1503.02584} {\path{arXiv:1503.02584}},
  \href {https://doi.org/10.1088/0004-637X/807/1/50}
  {\path{doi:10.1088/0004-637X/807/1/50}}.

\bibitem{koposov2015}
S.~E. Koposov, V.~Belokurov, G.~Torrealba, N.~W. Evans, {Beasts of the Southern
  Wild: Discovery of nine Ultra Faint satellites in the vicinity of the
  Magellanic Clouds}, ApJ 805~(2) (2015) 130.
\newblock \href {http://arxiv.org/abs/1503.02079} {\path{arXiv:1503.02079}},
  \href {https://doi.org/10.1088/0004-637X/805/2/130}
  {\path{doi:10.1088/0004-637X/805/2/130}}.

\end{thebibliography}

\label{lastpage}
\end{document}